\documentclass[%
preprint,
amsmath,amssymb,
aps,
prfluids,
]{revtex4-2}

\usepackage{graphicx}
\usepackage{dcolumn}
\usepackage{bm}
\usepackage{gensymb}
\usepackage{textcomp}
\usepackage{todonotes}
\usepackage[colorlinks=true,citecolor=blue,linkcolor=blue]{hyperref}

\usepackage[normalem]{ulem}

\usepackage{cancel}
\usepackage{subeqnarray}
\usepackage{mathtools}

\def\be{\begin{equation}}
\def\ee{\end{equation}}

\usepackage{verbatim}

\usepackage{cleveref}

\crefformat{section}{\S#2#1#3} 
\crefformat{subsection}{\S#2#1#3}
\crefformat{subsubsection}{\S#2#1#3}

\def\ee{{\rm e}}
\def\ii{{\rm i}}

\begin{document}

\preprint{APS/123-QED}

\title{Modified Stokes drift due to surface waves and corrugated sea-floor interactions with and without a mean current}

\author{Akanksha Gupta}
\affiliation{Department of Mechanical Engineering,
Indian Institute of Technology, Kanpur, U.P. 208016, India.}
\altaffiliation[Also at ]{
Aix Marseille University, CNRS, Centrale Marseille, IRPHE (UMR 7342), Marseille 13384, France.}

\author{Anirban Guha}%
\email{anirbanguha.ubc@gmail.com}

\affiliation{School of Science and Engineering, University of Dundee, Dundee DD1 4HN, UK.}




\date{\today}

\begin{abstract} 

In this paper, we show that Stokes drift may be significantly affected when an incident intermediate or shallow water surface wave travels over a corrugated sea-floor. The underlying mechanism is Bragg resonance -- reflected waves generated via nonlinear resonant interactions between an incident wave and a rippled bottom.  We theoretically explain the fundamental effect of two counter-propagating Stokes waves on Stokes drift and then perform numerical simulations of Bragg resonance using High-order Spectral method. A monochromatic incident wave on interaction with a patch of bottom ripple yields a complex interference between the incident and reflected waves.  When the velocity induced by the reflected waves exceeds that of the incident, particle trajectories reverse, leading to a backward drift. Lagrangian and Lagrangian-mean trajectories reveal that surface particles near the up-wave side of the patch are either trapped or reflected, implying that the rippled patch acts as a non-surface-invasive particle trap or reflector. On increasing the length and amplitude of the rippled patch; reflection, and thus the effectiveness of the patch, increases. The inclusion of realistic constant current shows noticeable differences between Lagrangian-mean trajectories with and without the rippled patch. Theoretical analysis reveals additional terms in the Stokes drift arising from the particular solution due to 
mean-current--bottom-ripple interactions, irrespective of whether Bragg resonance condition is met. Our analyses may be useful for designing artificial, corrugated sea-floor patches for mitigating microplastics and other forms of ocean pollution. We also expect that sea-floor corrugations, especially in the nearshore region, may significantly affect oceanic tracer transport.

\end{abstract}

\maketitle

\section{Introduction}
\label{sec:Intro}
Surface gravity waves existing at the ocean free surface cause floating particles, in addition to its periodic to-and-fro motion, a net transport in the direction of wave propagation, commonly known as the \emph{Stokes drift} \citep{stokes1847}. While periodic motion of fluid parcels arise from the  velocity field set by  surface waves at the leading order,
Stokes drift is a consequence of the fact that fluid parcels spend more time in the forward-moving region under the crest than the backward-moving region under the trough \citep{van2017stokes}. \citet{kenyon1969stokes} evaluated
Stokes drift for random surface waves in terms of the directional energy spectrum obtained empirically for fully developed seas (Pierson-Moskowitz). Stokes drift can be inferred from high-frequency radar, has the potential to be estimated from satellite measurements, and has already been accurately measured in the laboratory \citep[and references therein]{Sebille2020}. 

Even though Stokes drift is proportional to the square of wave steepness (therefore a small quantity compared to the leading order orbital motions), its magnitude is still significant and has important consequences \citep{mcwilliams1999wave}. It is known to play a key role in sediment transport \citep{kumar2017effect} and the local modeling of marine oil spills \citep{clark2015quantification}. Furthermore, it significantly contributes to the trajectories of drifters, especially for search and recovery missions (for example Airline MH370 crash \cite{trinanes2016analysis}), and is also important in mapping  the  pathways  of  plastic  pollution  including  microplastics in the global oceans \citep[and references therein]{van2017stokes,Sebille2020}. Stokes drift is considered to be a crucial player in the cross-shelf exchange of nearshore tracers like pathogens, contaminants, nutrients, larvae and  sediment \citep{kumar2017effect}.

Studies on Stokes drift have primarily approximated the ocean bottom to be a flat surface, while in reality, ocean bottom topography is spatially varying. Seafloor variations, much like mirrors and lenses in optics, can significantly affect the propagation of incident waves \citep{elandt2014surface}. An incident surface wave can interact with small undulations at the ocean bottom (hereafter, bottom ripple), giving rise to a reflected surface wave. This special type of wave-triad interaction is known as \emph{Bragg resonance}; here, the bottom ripple acts as a stationary wave and mediates energy transfer between the incident and reflected waves  \citep{davies1982reflection,mei1985resonant,heathershaw1982seabed}.
Resonant reflection is maximum when the wavenumber of the bottom corrugations is approximately twice the wavenumber of the incident surface wave. Bragg resonance strongly affects the wave spectrum in continental shelves and coastal regions \citep{Ball}, and also modifies the shore-parallel sandbars \citep{heathershaw1985resonant,elgar2003bragg}. Indeed there are many examples of Bragg phenomena in natural settings, e.g.\, Rotterdam waterway, Cape Cod Bay in Massachusetts, and near numerous shorelines \citep[and references therein]{reza_fabry}.

Bragg reflection causes standing or partially standing waves in the up-wave side of the sandbars, resulting in sediment transport and erosion or growth of the bars \citep{heathershaw1982seabed, elgar2003bragg}. On the up-wave side of the bars, the reflection coefficient is more or less constant; it rises to its peak value upto a short distance before falling, which leads to areas of preferential erosion and sediment transport \citep{heathershaw1982seabed}. Hence one can intuitively infer that in the vicinity of a bottom ripple (especially on the up-wave side), Stokes drift, and therefore surface particle transport, might be severely affected. To the best of our knowledge, the effect of sea-floor corrugations are not factored in while calculating Stokes drift. 

\begin{figure}
    \centering
    \includegraphics[width=\textwidth]{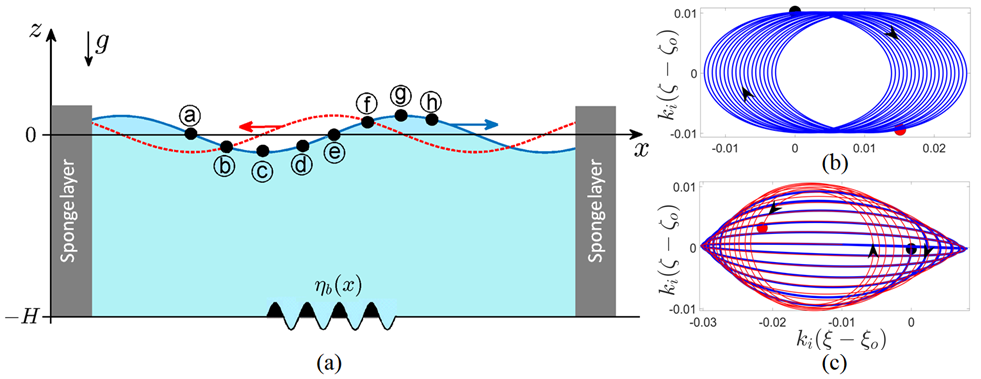}
    \caption{(a) Schematic of surface wave and bottom ripple with particles at different positions (labeled \textcircled{a}--\textcircled{h}) at $t=0$. Interaction of the rightward propagating incident wave (deep blue curve centered at $z=0$) with the bottom ripple (undulations centered at $z=-H$) leads to the generation of a leftward propagating reflected wave (red dashed curve centered at $z=0$). In reality, many reflected waves are generated from a finite patch, but here just one is shown for simplicity. (b) Trajectory of a particle in the absence of bottom ripple, and (c) trajectory of particle labeled \textcircled{d} in the presence of bottom ripple. In (b)--(c), black and red filled circles respectively denote initial and final location of the particle. Blue curves represent clockwise (forward) trajectories while red curves represent anti-clockwise (backward) trajectories. The axes in  (b)--(c) denote non-dimensional displacements from the initial position ($k_i$ is the wavenumber of the incident wave).}
   \label{fig:Schematic_diagram}
\end{figure}

The objective of this paper is to theoretically and numerically understand in a two-dimensional setting whether Stokes drift is significantly affected by undulations on the sea-floor, see Fig.\, \ref{fig:Schematic_diagram}(a) for a schematic diagram.  While details would follow in the rest of the paper, representative results are given in Figs.\, \ref{fig:Schematic_diagram}(b)--\ref{fig:Schematic_diagram}(c). Figure \ref{fig:Schematic_diagram}(b) shows the standard Stokes drift (in the absence of bottom ripple, the net drift, as expected, is forward), while Fig.\, \ref{fig:Schematic_diagram}(c) reveals that  Stokes drift in the presence of bottom ripple changes sign from positive to negative (i.e., undergoes backward motion). Stated otherwise, the presence of a bottom ripple can cause a large cancellation of the unidirectional drift, and thus cause \emph{particle trapping} in its vicinity.

The outline of the paper is as follows. In Sec.\,\ref{sec:2}, 
we briefly introduce the classical results of Stokes wave theory and also consider a special case of two counter-propagating waves and their resultant Stokes drift in finite, but constant water depth. Theoretical foundations of Bragg resonance, the High-order Spectral method, and the effect of Bragg resonance on the trajectories of particles at different initial locations  are discussed in Sec.\,\ref{sec:bragg_no_current}. The effects of the length of the rippled bottom patch and its amplitude on particle trajectories are discussed in Sec.\,\ref{sec:diff_U_ab}, and the analysis is extended to include the effects of constant mean current in Sec.\,\ref{sec:with_current}. We summarize and conclude the paper in Sec.\,\ref{summary}.

\section{Particle trajectories and Stokes drift in a fluid of constant  depth}
\label{sec:2}

\subsection{The classical Stokes wave solution}

We consider an incompressible, inviscid and irrotational  fluid inside the domain $\mathcal{B}=\{(x,z)\in \mathbb{R}^2: -H+\eta_b(x)\!<\!z\!<\!\eta(x,t)\}$, where $H$ is the mean fluid depth, $\eta$ denotes the free surface elevation, and $\eta_b$ denotes the rippled bottom  profile; see Fig.\, \ref{fig:Schematic_diagram}(a). We also assume a constant background current $U_0$ in the streamwise direction. Within the fluid domain $\mathcal{B}$, the  velocity potential $\phi(x,z,t)$ satisfies the governing equation (GE)
\begin{equation}
\phi_{,xx}+\phi_{,zz}=0,
\label{eq:Gov_eq}
\end{equation}
where comma in the subscript denotes partial derivative {($\phi_{,x}={\partial \phi}/{\partial x}$)}.
Impenetrability condition  {(ImC)} holds at the bottom boundary,  $z=-H+\eta_b(x)$:
 \begin{equation}
   \phi_{,z}-(\phi_{,x}+U_0) \eta_{b,x}=0,
   \label{eq:kin1_modified}
\end{equation}
which, in case of a flat bottom, simply leads to 
\begin{equation}
   \phi_{,z}=0
   \label{eq:kin1}
\end{equation}
at $z=-H$. The free surface at $z=\eta(x,t)$ evolves in time through the kinematic {(KBC)} and dynamic boundary conditions {(DBC)}, which can be respectively written as
\begin{subequations}
\begin{align}
&  \eta_{,t}+(\phi_{,x}+U_0)\eta_{,x}-\phi_{,z}=0, \label{eq:kin2}\\
&  \phi_{,t}+\frac{1}{2}[(\phi_{,x})^2+(\phi_{,z})^2]+U_0 \phi_{,x}+g\eta=0\label{eq:dyn},
\end{align}
\end{subequations}
where $g$ denotes gravitational acceleration. The spatio-temporal evolution of periodic, weakly nonlinear water waves can be obtained by solving the governing equation Eq. \eqref{eq:Gov_eq} and the relevant boundary conditions using perturbation expansions (with wave steepness `$ka$' taken as the small parameter, where  $a$ is wave's amplitude, and $k\in \mathbb{R}^+$ is its wavenumber). For example, the expressions for $\eta$ and $\phi$ for weakly nonlinear surface waves over fluid of arbitrary (but constant) depth $H$ up to second-order (i.e., $\mathcal{O}((ka)^2)$)  in the absence of background current is given by the classical Stokes theory 
\begin{subequations}
\begin{align}
\eta(x,t)&=a\cos(kx-\omega t)+ka^2\bigg[\dfrac{3-\tanh^2(kH)}{4\tanh^3(kH)}\bigg]\cos(2kx-2\omega t)+\mathcal{O}((ka)^3),\label{eq:Stokes_eta}\\
\phi(x,z,t)&=\dfrac{a\omega}{k\sinh(kH)}\bigg\{ \cosh[k(z+H)] \sin(kx-\omega t) + \nonumber \\
& 
ka\dfrac{3\cosh[2k(z+H)]}{8\sinh^3(kH)}\sin(2kx- 2\omega t)\bigg\}-(ka)^2\dfrac{1}{2\sinh(2kH)}\dfrac{gt}{k}+\mathcal{O}((ka)^3),\label{eq:Stokes_phi}
\end{align}
\end{subequations}
where $\omega\in \mathbb{R}$ denotes frequency. In the presence of a constant current $U_0$, the velocity potential $\phi(x,z,t)$ is simply the RHS of Eq. \eqref{eq:Stokes_phi} with $U_0x$ added, which is correct up to a constant. Stokes second-order theory reveals the steepening of surface wave crests and flattening of troughs for wave propagation over a flat bottom. However, there is yet another second-order motion that can arise; this is due to the interactions between first-order motion at the free surface and first-order undulations on the sea-bed \citep{heathershaw1985resonant}. The latter is known as Bragg reflection -- an incident surface wave resonantly interacts with the bottom ripple of twice its wavenumber to generate a  reflected wave. This aspect will be considered in Sec. \ref{sec:bragg_no_current}.

\subsection{Stokes drift and particle trajectories for two counter-propagating Stokes waves}
\label{counter_waves}

Let us consider two counter-propagating Stokes waves in the absence of background current, the wave with surface elevation $\eta_1(x,t)$ travels to the right while that with $\eta_2(x,t)$ propagates to the left:
\begin{subequations}
\begin{align}
&\eta_1= a^{(1)}  \,  \cos(kx-\omega t) + a^{(2)}   \,  \cos(2kx-2\omega t), \label{eta1_new} \\
&\eta_2= a^{(1)}  R \, \cos(kx+\omega t+\theta) +  a^{(2)}  R^2 \,  \cos(2kx+2\omega t+\theta),
\label{eta2_new}
\end{align}
\end{subequations}
where $\omega>0$, $a^{(1)} \equiv a$, $a^{(2)} \equiv ka^2[3-\tanh^2(kH)]/[4\tanh^3(kH)]$, $R\geq 0$ is a parameter and $\theta$ is the phase difference between the waves. Furthermore, the wave amplitudes must be small compared to both the wavelength and the water depth for each wave component, which stated mathematically yields \citep{kenyon1969stokes}:
$$
a^{(1),(2)} /H\ll 1\,\,\mathrm{and}\,\,k a^{(1),(2)}\ll 1.
$$
The potential due to the superposed Stokes wave $\eta=\eta_1+\eta_2$ is given by 
\begin{equation}
    \phi=\phi_1+\phi_2,
    \label{eq:total_phi}
\end{equation}
where $\phi_1$ is simply the expression in Eq. \eqref{eq:Stokes_phi}, and $\phi_2$ can be obtained by replacing $a$ by $Ra$ and $\omega$ by $-\omega$ in Eq. \eqref{eq:Stokes_phi}. The velocity field  can be straightforwardly obtained from the velocity potential: $\mathbf{u}\equiv (u,w)=(\phi_{,x},\phi_{,z})$.

Our primary objective is to obtain the trajectory $(x^{(p)}(t),z^{(p)}(t))$  of a certain tracer particle with initial position $(x_0,z_0)$. It is more convenient to represent displacements from the initial position: $\boldsymbol{\xi} \equiv \left(\xi(t),\zeta(t)\right)\equiv \left(x^{(p)}(t)-x_0,z^{(p)}(t)-z_0\right)$. The trajectory can be obtained from the path-line equation:
\renewcommand{\theequation}{\arabic{section}.\arabic{equation}a,b}
\begin{equation}
    \frac{d \xi}{d t}=u(x^{(p)},z^{(p)},t), \quad \frac{d \zeta}{d t}=w(x^{(p)},z^{(p)},t).
    \label{eq:pathline}
\end{equation}
\renewcommand{\theequation}{\arabic{section}.\arabic{equation}}

The assumption of small excursions from the initial position allows the application of the classical approximation theory \citep{kundu20fluid,constantin2008particle}, where $u(x^{(p)},z^{(p)},t)$ and $w(x^{(p)},z^{(p)},t)$ are Taylor expanded about $(x_0,z_0)$. This yields
\begin{subequations}
\begin{align}
u(x^{(p)},z^{(p)},t)&= u(x_o,z_o,t)+\xi\, u_{,x}|_{(x_o,z_o)} + \zeta\,u_{,z}|_{(x_o,z_o)} + .... \label{eq:u_tayl} \\
w(x^{(p)},z^{(p)},t)&= w(x_o,z_o,t)+\xi\, w_{,x}|_{(x_o,z_o)} + \zeta \,w_{,z}|_{(x_o,z_o)} + ....  \label{eq:w_tayl}
\end{align}
\label{eq:taylor_exp}
\end{subequations}
The variables $\boldsymbol{u}$ and $\boldsymbol{\xi}$ can be expanded as a perturbation series
\begin{subequations}
\begin{align*}
\boldsymbol{u}(x_o,z_o,t)&= \boldsymbol{u}^{(1)}(x_o,z_o,t)+ \boldsymbol{u}^{(2)}(x_o,z_o,t)+\mathcal{O}(\epsilon^3),\\
\boldsymbol{\xi}(x_o,z_o,t)&= \boldsymbol{\xi}^{(1)}(x_o,z_o,t)+ \boldsymbol{\xi}^{(2)}(x_o,z_o,t)+\mathcal{O}(\epsilon^3).
\end{align*}
\end{subequations}
Here superscript `$(n)$' denotes $\mathcal{O}(\epsilon^n)$,
where the small parameter $\epsilon$  scales as wave's steepness $ka$.

\subsubsection{Linear motion:}

Velocity field at $\mathcal{O}(\epsilon)$ can be calculated from the linear part of the velocity potential and then using Taylor expansion (Eq. \eqref{eq:taylor_exp}):
\begin{subequations}
\begin{align}
    u^{(1)}&=\dfrac{a\omega \cosh[k(z_o+H)]}{\sinh(kH)}  \bigg\{ \cos(kx_o-\omega t) -  R \, \cos(kx_o+\omega t+\theta) \bigg\}, \label{eq:u_order_1} \\
    w^{(1)}&=\dfrac{a\omega \sinh[k(z_o+H)]}{\sinh(kH)}  \bigg\{\sin(kx_o-\omega t) -  R \, \sin(kx_o+\omega t+\theta)\bigg \}.
    \label{eq:w_order_1}
\end{align}
\end{subequations}
The corresponding linear displacements are obtained from   Eq. \eqref{eq:pathline}:
\begin{subequations}
\begin{align}
& \xi^{(1)}=-a \frac{\cosh[k(z_o+H)]}{\sinh(kH)} \bigg\{\sin(kx_o-\omega t) +  R \, \sin(kx_o+\omega t+\theta)\bigg\}, \label{x-para}\\
& \zeta^{(1)}=a \frac{\sinh[k(z_o+H)]}{\sinh(kH)}\bigg \{  \cos(kx_o-\omega t) + R \, \cos(kx_o+\omega t+\theta)\bigg\}. \label{z-para}
\end{align}
\end{subequations}
For $R=0$, we recover the standard parameterization given in \citet{kundu20fluid}; it yields the parametric representation of elliptical trajectory (for shallow and intermediate water region) for  linear water waves. While Eqs. \eqref{x-para}--\eqref{z-para}  is not a surprising outcome, it nevertheless reveals an  interesting result; for $R=1$, the ellipse collapses into a line, which is given by 
\begin{equation}
    \zeta=-\dfrac{\tanh[k(z_o+H)]}{\tan(kx_o+\theta/2)} \xi. \label{eq:linear_trajectory}
\end{equation}
When $kx_o+\theta/2=0$, the trajectory is vertical while for $kx_o+\theta/2=\pm \pi/2$,  we obtain a horizontal trajectory. Such linear trajectories are due to the standing wave pattern generated by the counter-propagating surface waves.

\subsubsection{Stokes Drift:}
\label{subsec:stokes_drift}
Stokes drift results from the quadratic ($\mathcal{O}(\epsilon^2)$) terms arising from the product of two linear ($\mathcal{O}(\epsilon)$) terms in Eqs. \eqref{eq:u_tayl}--\eqref{eq:w_tayl}:  
\begin{equation}
   \begin{bmatrix}
u_{SD} & w_{SD} 
\end{bmatrix}
=
\begin{bmatrix}
\xi^{(1)} &  \zeta^{(1)} 
\end{bmatrix}
\left.
 \begin{bmatrix}
 u_{,x}^{(1)}  & w_{,x}^{(1)} \\ \\ 
u_{,z}^{(1)}   & w_{,z}^{(1)} 
\end{bmatrix} \right  \vert_{(x_0,z_0)}, 
\label{eq:Stokes_drift_formula}
\end{equation}
which can be written more compactly as 
$\boldsymbol{u_{SD}} =\boldsymbol{\xi^{(1)}} \cdot \nabla \boldsymbol{u^{(1)}}|_{(x_0,z_0)}$. Note that Stokes drift $\langle\boldsymbol{u_{SD}}\rangle$ is obtained after time averaging (denoted by $\langle\, \rangle$) the velocity  $\boldsymbol{u_{SD}}$ over one time-period.  

For two counter-propagating Stokes waves $\eta_1$ and $\eta_2$, we obtain
\begin{subequations}
\begin{align}
&u_{SD}=\frac{a^2 \omega k}{\sinh^2(kH)} \Big\{\cosh^2[k(z_o+H)]\sin^2(kx_o-\omega t)+\sinh^2[k(z_o+H)]\cos^2(kx_o-\omega t) \Big\}- \nonumber \\&
\frac{a^2 \omega k R^2}{\sinh^2(kH)} \Big\{\cosh^2[k(z_o+H)]\sin^2(kx_o+\omega t+\theta)+\sinh^2[k(z_o+H)]\cos^2(kx_o+\omega t+\theta) \Big\}, \label{SD_1} 
\\
&w_{SD}= -a^2  \omega k R \frac{\sinh[2k(z_o+H)]}{\sinh^2(kH)}\sin(2\omega t+ \theta). \label{SD_2} 
\end{align}
\end{subequations}
The \emph{modified} Stokes drift can now be straight-forwardly obtained as follows:
\begin{subequations}
\begin{align}
\langle{u_{SD}}\rangle &= a^2 \frac{\omega k(1-R^2)}{2\sinh^2{(kH)}}\cosh[2k(z_o+H)],\label{z_SDa} \\
\langle{w_{SD}}\rangle & =0.\label{z_SDb}
\end{align}
\end{subequations}
Substitution of $R=0$ results in the standard Stokes drift in a fluid of arbitrary (constant) depth, first obtained by \citet{ursell1953long}. 

\begin{figure}
    \centering
    \includegraphics[width=\textwidth]{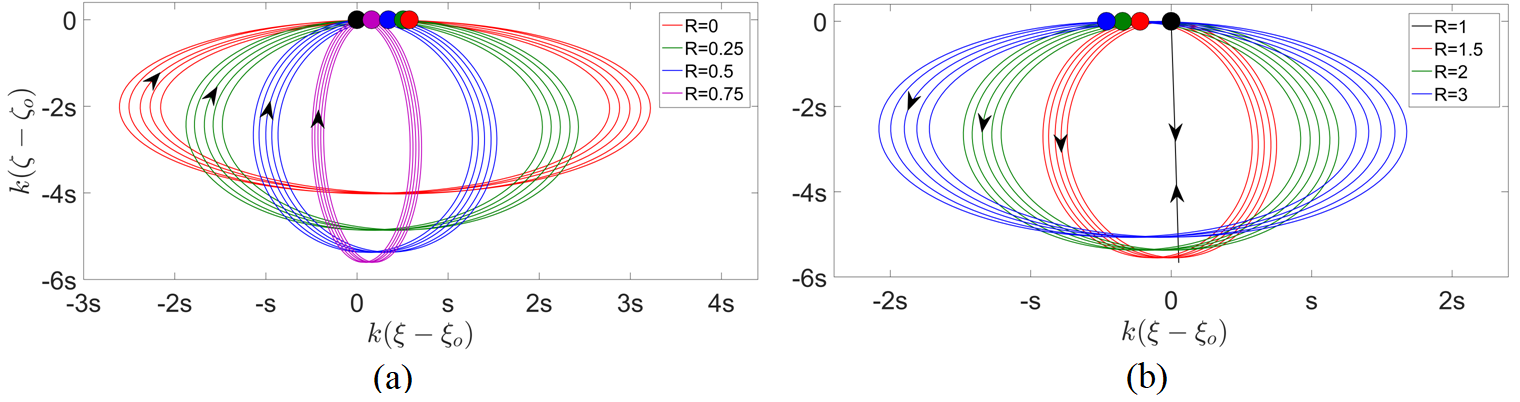}
    \caption{Particle trajectories resulting from two counter-propagating Stokes waves for $\theta=0$ and different values of $R$: (a)  $R<1$, and (b) $R \ge 1$. 
    Black filled circle denotes initial position, while colored filled circles (with color corresponding to that of the respective trajectory) denote position after five time-periods. 
    The quantity $s=0.005$ for the parameters chosen.}
    \label{fig:variation_with_R_new}
\end{figure}

Equations  \eqref{z_SDa}--\eqref{z_SDb} reveal that the Stokes drift due to two counter-propagating waves is the superposition of the drift due to the individual waves. This fact for a random wave field was already mentioned by \citet{kenyon1969stokes}, however the problem with two counter-propagating waves being simpler reveals certain important properties.
The first thing to observe is that the modified Stokes drift is \emph{independent} of $\theta$. To understand the dependence on $R$, we plot the particle trajectories in Fig.\, \ref{fig:variation_with_R_new} for various $R$ values, keeping the energy $a^2(1+R^2)=\mathrm{constant}$. For $R \in [0,1)$, the drift is in the positive direction and the particle trajectory is clockwise, see Fig.\, \ref{fig:variation_with_R_new}(a).
The drift velocity \emph{decreases} with increasing $R$ (as obvious from Eq. \eqref{z_SDa}); moreover, as $R$ increases, the horizontal span (i.e., the horizontal distance a particle travels over one time-period) decreases while the vertical span increases. Finally, the horizontal span, and therefore the Stokes drift, becomes zero at  $R=1$, see Fig.\, \ref{fig:variation_with_R_new}(b) where $R\ge 1$ is plotted. For $R=1$ case, the particle oscillates along a linear trajectory (vertical in this case) given by Eq. \eqref{eq:linear_trajectory}. For $R>1$, the drift again increases but in the negative direction, and the particle trajectory this time is counter-clockwise,  which is also evident from  Eq. \eqref{z_SDa}. As $R$ increases, there is an increase in the horizontal and decrease in the vertical span. 
We also observe in Figs.\, \ref{fig:variation_with_R_new}(a)--\ref{fig:variation_with_R_new}(b) that as $R$ increases, the final position moves more in the backward direction.

In summary, this section has revealed in a straight-forward way that two counter-propagating Stokes waves in a fluid of constant depth (flat bottom) strongly alters the forward drift, which becomes zero when the waves have the same amplitude. The latter is due to the standing wave pattern generated by the counter-propagating Stokes waves. The drift again increases, and its direction reverses when the leftward traveling wave has a higher amplitude. Since Stokes drift is a time-averaged quantity (over a time-period), the phase difference between the waves does not affect the drift velocity. While there could be numerous situations where one can obtain counter-propagating waves  (e.g., those formed by shore-ward propagating waves and waves reflected from the shore), the situation which we will focus next, and is central to this paper, is Bragg reflection, which arises under certain conditions when an incident surface wave propagates over a bottom ripple.

\section{Bragg resonance and particle trajectories in the absence of a background current}
\label{sec:bragg_no_current}

{\subsection{Bragg resonance condition and amplitude evolution}}
Nonlinear interaction of an incident surface wave with  bottom ripples gives rise to a resonant wave, which satisfies the following triad  condition \citep{liu1998generalized}:
\begin{equation*}
   k_i+ k_r = k_b,\quad  \omega_i + \omega_r = \omega_b.
\end{equation*}
Here the subscripts $i$, $r$, and $b$ respectively denote the incident wave, the reflected wave, and the bottom ripples. Unlike the classical wave triad formed by three waves, one of the waves is replaced by a stationary bottom ripples in the case of a Bragg resonance. Since the bottom ripples are stationary ($\omega_b=0$), it implies $\omega_r=-\omega_i$, i.e.\, the reflected wave travels in a direction opposite to the incident wave with the same magnitude. The dispersion relation for surface waves, 
\begin{equation}
\omega^2-gk\tanh{(kH)}=0,
\label{eq:disp_rel}
\end{equation}
yields $k_r=k_i$, and hence $k_b=2k_i$.

To derive the amplitude evolution equations for Bragg resonance, we  represent the constituent surface waves in the form $a_i(\tau)\exp{[\ii(k_ix-\omega_it)]}+\mathrm{c.c.}$ and $a_r(\tau)\exp{[\ii(k_rx-\omega_rt)]}+\mathrm{c.c.}$, where $\mathrm{c.c.}$ denotes complex conjugate, $\tau=\epsilon t$ is the slow time scale, and $\epsilon$ is a small parameter.
Assuming bottom ripples centered around $z=-H$ of the form $\eta_b(x)=a_b \sin(k_b x)$   where $a_b$ is the ripple's amplitude, the wave amplitude evolution equations for Bragg resonance are given by (see Appendix \ref{Sec:amp_evol} for the derivation):
\renewcommand{\theequation}{\arabic{section}.\arabic{equation}a,b}
\begin{align}
\frac{da_i}{d\tau}= \ii \lambda \frac{\omega_i}{2g} a_b\bar{a}_r\qquad;\qquad\frac{da_r}{d\tau}=\ii \lambda \frac{\omega_r}{2g} a_b\bar{a}_i,\label{eq:Triad}
\end{align}
\renewcommand{\theequation}{\arabic{section}.\arabic{equation}}

\noindent where  $\lambda=\omega_i\omega_r[\sinh{(k_iH)}\sinh{(k_rH)}]^{-1}$,  and overbar denotes complex conjugate. Equation (\ref{eq:Triad}) does reveal via the coefficient $\lambda$ that the effect of Bragg resonance is minimal in the deep water limit and increases as the water gets shallower. We note here that although (\ref{eq:Triad}) is simpler and more analytically tractable than the situation with a finite patch of bottom ripple (see Refs.\, \cite{davies1982reflection,davies1984surface,mei1985resonant} for detailed theoretical analysis), it does capture the essential physics of reflected wave generation due to Bragg resonance.

{\subsection{Numerical scheme for simulating Bragg resonance}}
\label{sec:particle_traj}

We intend to numerically simulate the configuration shown in Fig.\, \ref{fig:Schematic_diagram}(a), i.e., Bragg reflection for a finite patch of bottom ripple. The horizontal domain extends from $0$ to $2\pi$, and the bottom topography at $z=-H$ is given by the following equation:
\begin{equation} \label{eq:baines_profile}
\eta_b(x) = \Bigg\{ \begin{array}{l@{\quad}l} a_b \sin(k_b x)  & \textrm{, $\pi-l \le x \le \pi+l$}, \\ 
0 & \textrm{, otherwise},\\ \end{array} 
\end{equation}
where $l$ is the half-length of the patch. For numerical simulation, we employ an in-house code based on the High-order spectral (HOS) method, a highly accurate and efficient numerical method developed by \citet{dommermuth1987high} for studying wave propagation, wave-wave, and wave-topography interactions. 
Our code is capable of handling multi-layered flows (i.e., both surface and interfacial gravity waves) along with piece-wise constant vorticity in each layer as well.
The theoretical and numerical details have been comprehensively reported in  \citet{raj_guha_2019}, however for completeness, here we provide a brief overview. For the problem relevant to this work, we solve the Laplace equation Eq. \eqref{eq:Gov_eq} along with the boundary conditions Eq. \eqref{eq:kin1_modified}, Eqs. \eqref{eq:kin2}--\eqref{eq:dyn}
using HOS method, which can provide solutions up to an arbitrary order of nonlinearity.

In the HOS method the velocity potential at the free surface is described in terms of the surface potential
\begin{equation}
  \Phi^s(x,t)=\phi(x,\eta(x,t),t). 
\end{equation}
Thus the free surface boundary conditions Eq. \eqref{eq:kin2} and Eq. \eqref{eq:dyn} can be respectively expressed in terms of $\Phi^s$ as follows:
\begin{subequations} 
\begin{align}
\eta_{,t}&=-\eta_{,x}(\Phi_{,x}^s+U_0)+(1+\eta_{,x}^2)\phi_{,z},\label{eq:zakharov_eq1}\\
\Phi_{,t}^s&=-g\eta-\frac{1}{2}(\Phi_{,x}^s)^2+\frac{1}{2}(1+\eta_{,x}^2)\phi_{,z}^2-U_0\Phi_{,x}^s, \label{eq:zakharov_eq2}
\end{align}
\end{subequations}
\noindent which are often referred to as the Zakharov equations \citep{zakharov1968stability}. 

Note that for the time integration of Eqs. \eqref{eq:zakharov_eq1}--\eqref{eq:zakharov_eq2},  $\phi_z$ at each time-step is required. To this end we expand $\phi$ as a perturbation series up to a given order $n$:
\begin{equation*}
    \phi(x,z,t)=\sum_{m=1}^{n}\phi^{(m)}(x,z,t).    
\end{equation*}
At every order $m$, the velocity potential is further expressed as a sum of  Fourier basis function. Assuming solutions to be periodic in the $x$-direction, the solutions are represented as a discrete Fourier series: 
\begin{equation}
\phi^{(m)}= \sum_{n=-N}^{N-1}\left[A_n^{(m)}(t)\frac{\cosh{k_n(z+H)}}{\cosh{(k_n H)}} + B_n^{(m)}(t)\frac{\sinh{(k_n z)}}{\cosh{(k_n H)}} \right] \ee^{\ii k_nx}.
\label{Four1}
\end{equation}
However, direct substitution of Eq. \eqref{Four1} in the boundary conditions to obtain the unknown coefficients is inconvenient since $z$ has dependence on $x$ at the surface. The issue can be circumvented by expanding the surface potential as a Taylor series about $z=0$:
\begin{equation}
    \Phi^s(x,t)  =\sum_{m=1}^{n}\sum_{k=0}^{n-m}\frac{\eta^k}{k!}\frac{\partial^k }{\partial z^k}\phi^{(m)}\bigg\rvert_{z=0}.
    \label{eq:surf_poten_taylor}
\end{equation}
The above equation  can be expressed as a sequence of Dirichlet boundary conditions at each order $m$. The boundary conditions at each order depends on product of the  terms which have already been evaluated at the lower orders, therefore making the problem effectively linear at every order $m$. This is given as follows:
 \begin{equation}
    	\phi^{(m)}(x,0,t)=f_1^{(m)},\label{HOS_BC1}
    \end{equation}
    where
    \begin{subequations}
    \begin{align}
    	f_1^{(1)}&=\Phi^S,\\
        f_1^{(m)}&=-\sum_{k=1}^{m-1}\frac{\eta^k}{k!}\frac{\partial^k}{\partial z^k}\phi^{(m-k)}\bigg\rvert_{z=0}.
    \end{align}
    \end{subequations}

    Finally for the bottom boundary, the impenetrability condition yields:
    \begin{equation}
    \phi_{,z}^{(m)}(x,-H,t)=f_2^{(m)},\label{HOS_BC4}
\end{equation}
    where
        \begin{subequations}
   \begin{align}
    f_2^{(1)}&=U_0\eta_{b,x},\\
        f_2^{(m)}&=\sum_{k=1}^{m-1}\frac{\partial}{\partial x}\left[\frac{\eta_b^k}{k!}\frac{\partial^{k-1}}{\partial z^{k-1}}\phi_{,x}^{(m-k)}\bigg\rvert_{z=-H}\right].
    \end{align}
        \end{subequations}
    Using the boundary conditions, we obtain the  unknown coefficients $A_n$  and $B_n$ at every order $m$, leading to the full solution of $\phi^{(m)}$. At the next order $m+1$, the functions $f_1$ and $f_2$ can be evaluated by using the velocity potential and its derivative, which have been already evaluated at the previous order $m$.

The boundary value problem has been solved using $1024$ Fourier modes, and the order of nonlinearity ($n$) of the HOS method is kept at $2$. Amplification of round-off errors occurs at higher wavenumbers, which needs to be filtered out using a low-pass filter. For time integration fourth-order Runge-Kutta method has been implemented. Sponge layers have been used (see Fig.\, \ref{fig:Schematic_diagram}(a)) at the horizontal boundaries to prevent interactions between transmitted and reflected waves.

\vspace{0.5cm}

\subsection{Particles trajectories at various locations}
\label{subsec:part_traj}

At $t=0$, a rightward propagating monochromatic incident wave  $\eta=a_i\cos(k_i x)$, where  $a_i=0.01H$ and $k_iH=1$, is added to the free surface. The amplitude and wavenumber of the bottom ripple are respectively
$a_b=0.1H$ and $k_bH=2$. Hence the incident wave is an intermediate water surface wave that satisfies the Bragg resonance condition.
This is necessary since, as mentioned already, Bragg resonance would be minimal in the deep water limit. The bottom ripple profile given in Eq. \eqref{eq:baines_profile} is considered and the patch length is fixed at $2l=\pi/2$.

\begin{figure}
    \centering
    \includegraphics[width=\textwidth]{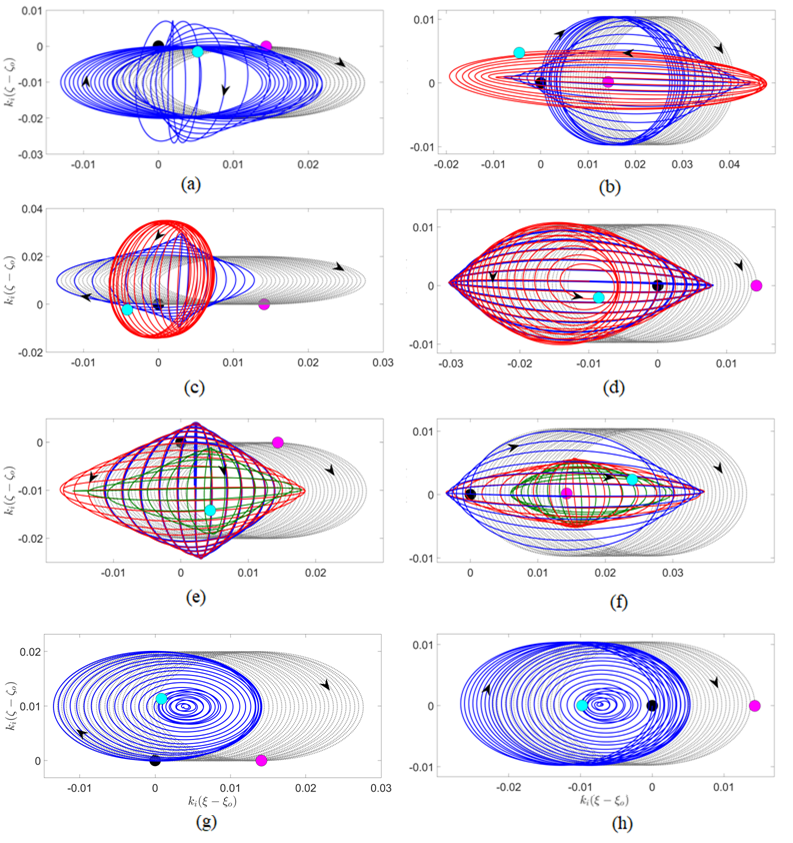}
    \caption{Trajectories of tracer particles at different initial locations, as shown in Fig.\, 1(a).  Blue colored trajectories show forward (clockwise) drift, red trajectories show backward (anti-clockwise) drift, while green trajectories show the next round of forward (clockwise) drift. Black and cyan filled circles respectively denote initial and final particle positions. Gray  trajectories denote  Stokes drift for a flat bottom topography. In this case the final position of the particle is represented by magenta filled circle. Note the initial position (denoted by  black filled circle) for the flat topography case is the same as the one with rippled bottom topography.}
    \label{fig:particle_trajectories}
\end{figure}

To understand the fate of particles at different initial locations, we place $8$ particles, labelled \textcircled{a}--\textcircled{h} on the free surface; see Fig.\, \ref{fig:Schematic_diagram}(a). The particles are evenly distributed in the domain $\pi/2 \! \leq \! x \! \leq \! 11\pi/8$ (they are at $\pi/8$ distance from each other), and their respective trajectories are depicted in Figs.\, \ref{fig:particle_trajectories}(a)--\ref{fig:particle_trajectories}(h). Note that particles \textcircled{c}--\textcircled{g} are initially above the bottom ripple, \textcircled{a}--\textcircled{b} are up-wave side while \textcircled{h} is down-wave side of the bottom ripples.
Fig.\, \ref{fig:particle_trajectories}(a) reveals that the parcel \textcircled{a} always undergo a forward drift (indicated by blue-colored trajectories denoting clockwise circulation), but the drift velocity decreases as soon as the reflected waves pass through it. The effect of the reflected wave becomes more pronounced on parcels \textcircled{b}--\textcircled{f}, and these parcels undergo anti-clockwise or backward drift (red-colored trajectories). The backward drift of parcels \textcircled{c}--\textcircled{f} is primarily due to multiple reflections from the finite rippled patch, and occurs when the velocity induced by the reflected wave exceeds that of the incident. Such negative drift can be readily observed if $R>1$ in Eq. \eqref{z_SDa}, or from Fig.\, \ref{fig:variation_with_R_new}(b).
The green trajectories for parcels \textcircled{e}--\textcircled{f} denote the next sequence of forward drift and occurs when the incident wave again dominates over the reflected. Parcels \textcircled{g}--\textcircled{h} undergo forward drift due to the transmitted wave, but their drift velocities decrease with time.

\begin{figure}
    \centering
    \includegraphics[width=\textwidth]{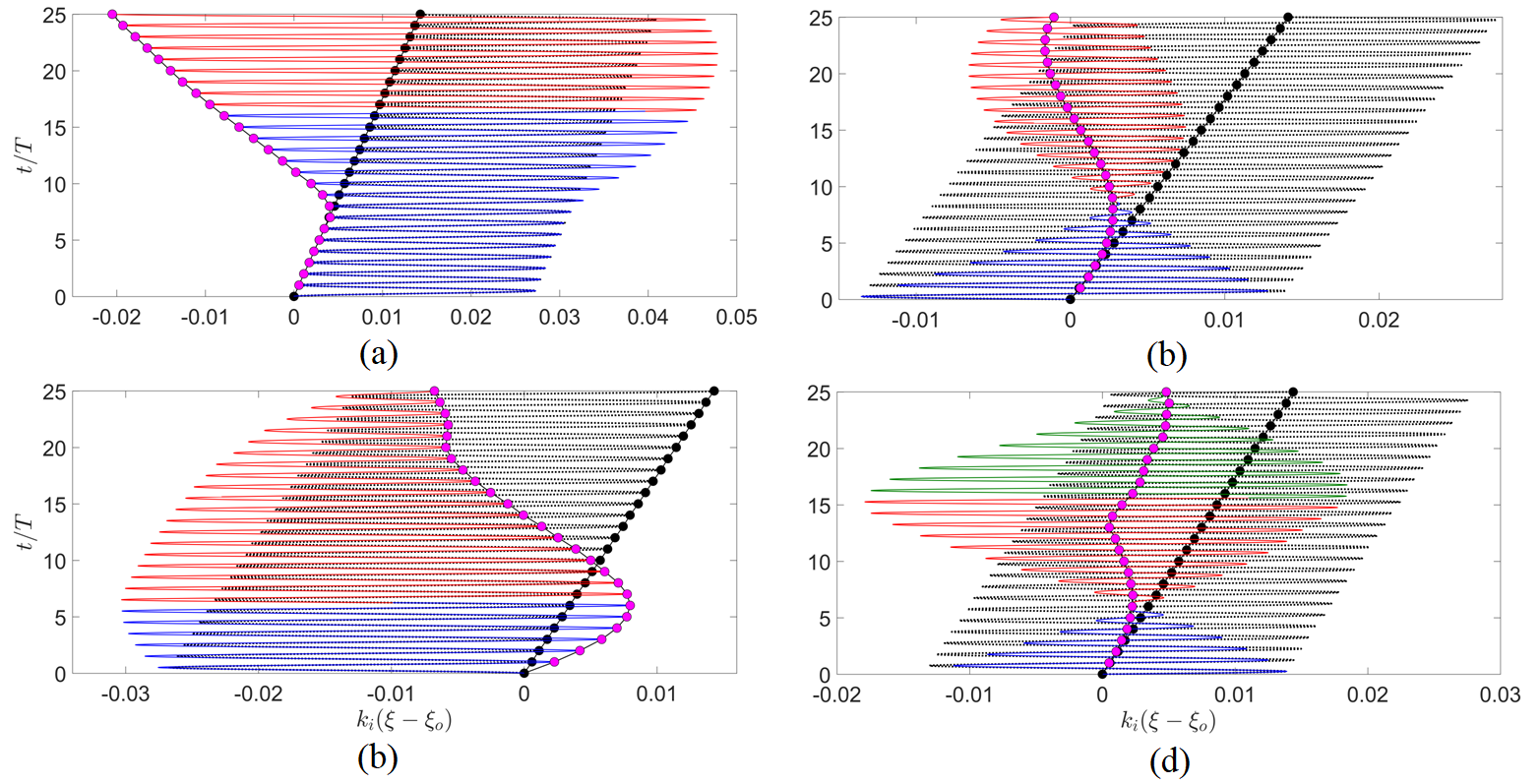}
    \caption{Horizontal particle trajectories versus time in the absence (small black dotted curves) and  presence  (red, blue and green curves with color scheme same as Fig.\, \ref{fig:particle_trajectories}) of the rippled bottom topography. Trajectories corresponding to the particles \textcircled{b}--\textcircled{e} are respectively shown in (a)--(d). Horizontal component of Lagrangian-mean trajectories without and with the bottom ripple are respectively represented by the large black and magenta dotted lines/curves; each dot denotes particle position after one time-period ($T$).}
    \label{fig:x_vs_t}
\end{figure}

Figure \ref{fig:x_vs_t} shows the variation of time-averaged horizontal transport  with time {(i.e., the $x$-component of the Lagrangian--mean trajectory)} for particles at  locations \textcircled{b}--\textcircled{e} for the problem discussed above. For flat bottom topography, the horizontal drift velocity is a positive constant (given by Eq. \eqref{z_SDa}  for $R=0$), leading to a linear growth of the horizontal distance (averaged over one time-period)  with time, shown by large black dotted line. The situation is different for the case with a rippled patch that causes Bragg resonance; the complex interference of incident and reflected waves lead to time-averaged trajectories  (denoted by large magenta dotted curves) which reveal strongly backward motion (Figs.\, \ref{fig:x_vs_t}(a)--\ref{fig:x_vs_t}(c)), or at least, minimal forward motion (Fig.\, \ref{fig:x_vs_t}(d)). Hence particles that are carried by the waves in the forward direction are non-invasively  ``reflected'' by the bottom ripple.

The simulations performed have been limited to $25$ time-periods so that the effect of sponge layers at the horizontal boundaries are minimal. If run for a longer time, the amplitudes of the incident and reflected waves would decay to zero. In a more realistic scenario, one can expect a continuous source of incident wave packets, interacting with the reflected waves that were generated in the past, as well as producing new reflected waves via Bragg resonance. The continuous generation of reflected waves would hinder the down-wave propagation of tracer particles. Hence we expect that the scenario depicted in our simulations would not be fundamentally different from the more realistic case.  

In summary, all parcels existing in the vicinity of the rippled patch experience a drastic change in their otherwise forward drift, the rippled patch acts as an effective non-surface-invasive \emph{particle trap} or a \emph{particle reflector}.


\section{Effect of bottom ripple's amplitude and patch size  on particle trajectory}
\label{sec:diff_U_ab}

The classical works of \citet{davies1982reflection} and \citet{davies1984surface} reveal that the reflection coefficient $C_R$ (and hence the amplitude of the reflected wave) increases on increasing the ripple amplitude $a_b$ or the patch length $2l$, succinctly given by the following equation:
\begin{equation}
    C_R=\dfrac{2a_bk_i}{2k_iH+\sinh{(2k_iH)}}\bigg[(-1)^m (2k_i/k_b)\dfrac{\sin\{m\pi (2k_i/k_b)\}}{(2k_i/k_b)^2-1}\bigg],
    \label{eq:Davies82}
\end{equation}
where $m\pi/k_b=l$, $m$ denotes the number of bottom ripples. For long waves ($k_iH\ll1$), Eq. \eqref{eq:Davies82} can be further simplified for resonant condition ($2k_i=k_b$) as follows:
\begin{equation*}
    C_R=\frac{a_b}{2H}\frac{m\pi}{2}.
\end{equation*}
This implies that  $a_b$ and $m$ (or $2l$) may have a significant effect on the modified Stokes drift, which we intend to study in this section.

\subsection{Variation of  bottom ripple's amplitude}

For this particular study we consider the bottom ripple profile Eq. \eqref{eq:baines_profile} with patch length fixed at $2l=\pi/2$ but ripple amplitude $a_b$ is varied.
We supply a continuous source of rightward propagating incident wave-packets with initial profile $\eta=0.005H[\tanh\{10(x-5\pi/12)\}-\tanh\{10(x-7\pi/12)\}] \cos(k_i x)$, where the peak wavenumber $k_iH=1$. The initial wave-packet profile is centered around $x=\pi/2$ and practically decays to zero before reaching $x=3\pi/4$, i.e. the starting point of bottom ripples. In other words, the bottom is flat in the `wave generation' region.

\begin{figure}
    \centering
    \includegraphics[width=0.7\textwidth]{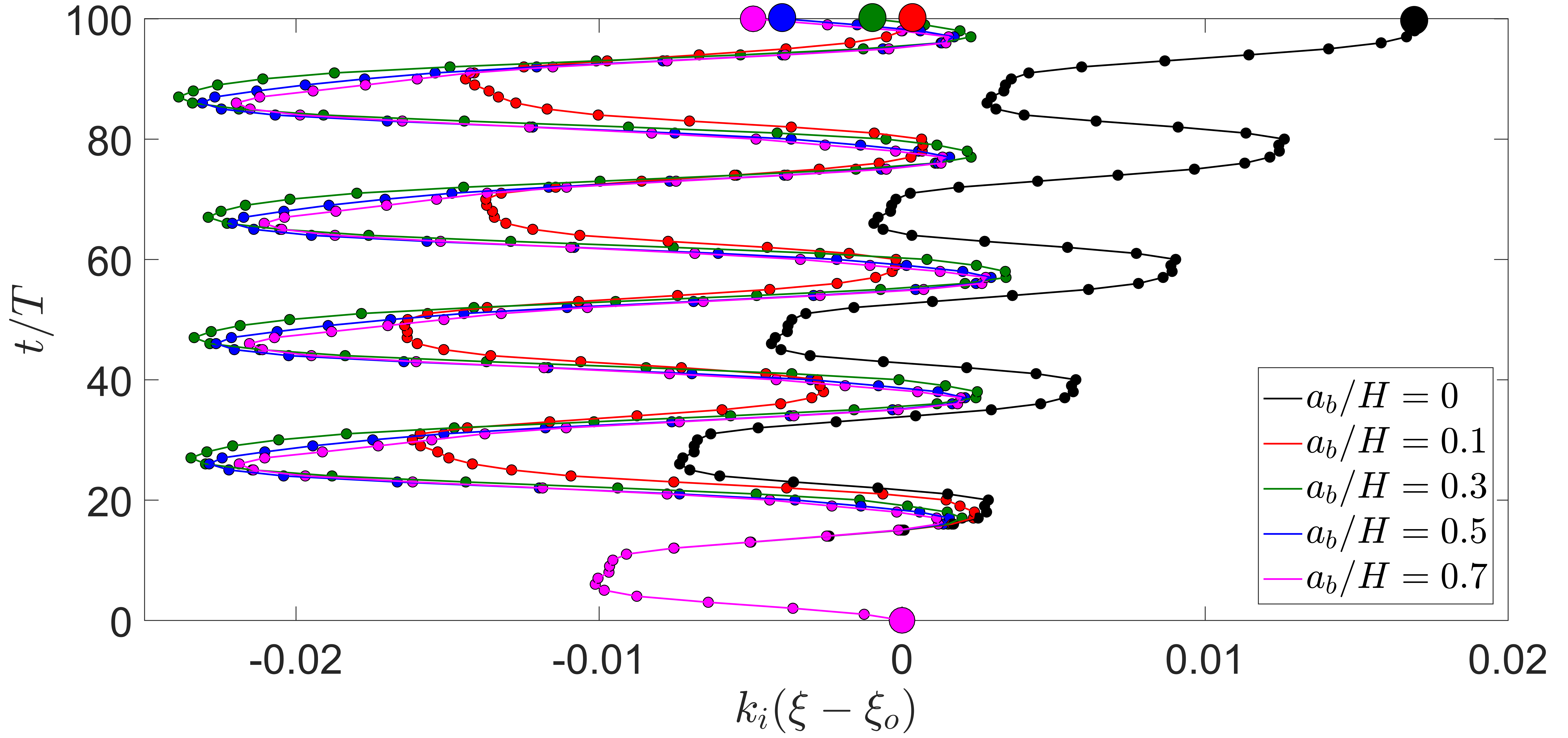}
    \caption{Horizontal component of the Lagrangian-mean trajectory of particle \textcircled{b} (initial abscissa: $x_0=5\pi/8$)  versus time for (i) flat bottom (black), (ii) $a_b/H=0.1$ (red), (iii) $a_b/H=0.3$ (green), (iv) $a_b/H=0.5$ (blue), and (v) $a_b/H=0.7$ (magenta). Simulations have been performed for a continuous source of incident wave-packets and patch length is held fixed at $2l=\pi/2$. Small dots represent position after one time-period, while large filled circles denote initial and final positions.
    }
    \label{fig:different_ab}
\end{figure}

Figure \ref{fig:different_ab} shows the variation of  the $x$-component of the Lagrangian-mean trajectory with time for particle \textcircled{b} (initial abscissa: $x_0=5\pi/8$) for $a_b/H=0,\,0.1,\,0.3,\,0.5$, and $0.7$. 
The simulation is run for $100T$ ($T$ denotes time-period corresponding to the peak wavenumber), and sponge layers are used at the horizontal  boundaries to absorb the reflected and transmitted waves.
For the flat bottom case  the particle, as expected, moves only in the forward direction with positive Stokes drift.  However as $a_b$ increases, the amplitude of the reflected waves also increase, leading to an increased backward drift (i.e.\, corresponds to $R>1$ for the simplified analysis in Eq. \eqref{z_SDa}). In summary, larger the amplitude of the bottom ripple, the more is its efficiency as a particle trap or reflector.

\subsection{Variation of  patch length}

\begin{figure}
    \centering
    \includegraphics[width=\textwidth]{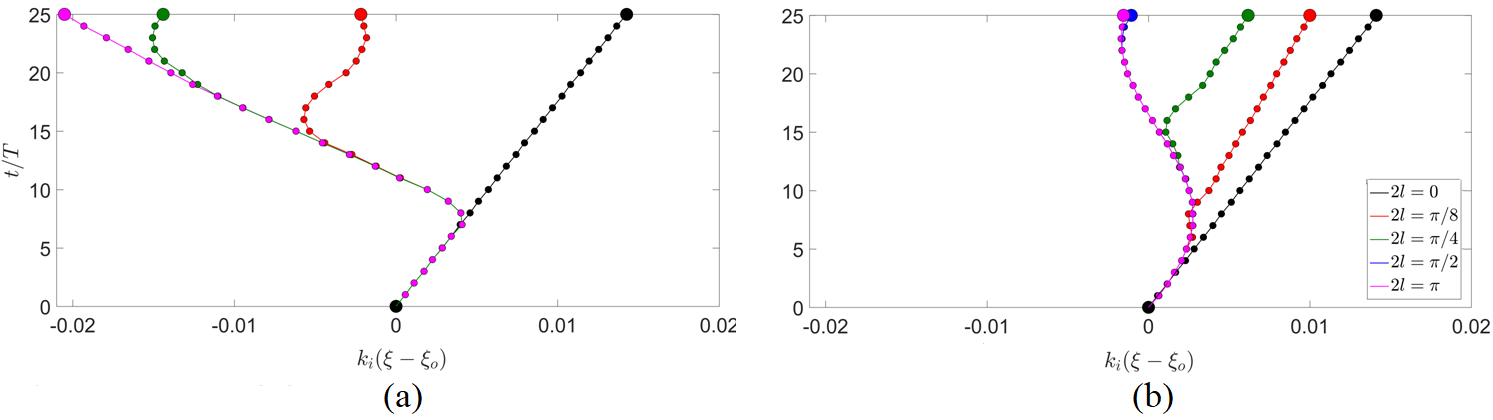}
    \caption{Horizontal component of the Lagrangian-mean trajectory of particle with initial position fixed at (a) \textcircled{b} (initial abscissa: $x_0=5\pi/8$), and (b) \textcircled{c} (initial abscissa: $x_0=3\pi/4$) for different patch sizes ($2l$): (i) flat bottom (black) ,  (ii) $\pi/8$ (red),  (iii) $\pi/4$ (green), (iv) $\pi/2$ (blue), and (v) $\pi$ (magenta). Simulations have been performed for a wave-train and bottom ripple's amplitude is held fixed at $a_b/H=0.1$. Small dots represent position after one time-period, while large filled circles denote initial and final positions.}
    \label{fig:patch_size_diff}
\end{figure}

By fixing the patch length in Sec. \ref{subsec:part_traj}, we observed that a particle's trajectory changes according to its initial position relative to the bottom ripples. Here we fix the position of a particle and study its trajectory for various patch lengths. 
Unlike Sec. \ref{subsec:part_traj}, where the patch was centered around $x=\pi$, here we fix the starting point of the patch at $x=3\pi/4$ and vary its end point:
\begin{equation} \label{eq:baines_profile_start_fixed}
\eta_b(x) = \Bigg\{ \begin{array}{l@{\quad}l} a_b \sin(k_b x)  & \textrm{, $3\pi/4 \le x \le 2l+ 3\pi/4$}, \\ 
0 & \textrm{, otherwise}.\\ \end{array} 
\end{equation}
The ripple amplitude is fixed at $a_b=0.1H$, and the free surface is  initialized with wave-trains as in Sec. \ref{subsec:part_traj}. We consider particle \textcircled{b} (initial abscissa at $x_0=5\pi/8$) and particle \textcircled{c} (initial abscissa at $x_0=3\pi/4$) and study for patch lengths $2l=0,\,\pi/8,\,\pi/4,\,\pi/2$ and $\pi$. Particle \textcircled{b} is initially situated at a distance $\pi/8$ up-wave of the first bottom ripple, and its   $x$-component of the Lagrangian-mean trajectory for various patch lengths is plotted in Fig.\, \ref{fig:patch_size_diff}(a). The same analysis is repeated for particle \textcircled{c} in Fig.\, \ref{fig:patch_size_diff}(b),  which is initially situated right above the starting point of the first bottom ripple. Both Figs.\, \ref{fig:patch_size_diff}(a)--\ref{fig:patch_size_diff}(b) show the obvious linear relation between the $x$-component of the Lagrangian-mean trajectory and time for the flat bottom case (since Stokes drift is a constant positive value). As the patch length is increased (recall that the ripple amplitude is small: $a_b=0.1H$), we observe a strong negative drift of particle \textcircled{b}, i.e.\, the particle is always reflected. However the reflection of particle \textcircled{c} is relatively weaker. Nevertheless for $2l\geq \pi/2$, particles in the vicinity of the up-wave side of the patch are \emph{always} either trapped or reflected.

The resonance effect becomes continuously sharper as the patch length increases, and theoretically tends to infinity for an infinite patch length. 
However, the theoretical analysis breaks down before this situation is reached \citep{davies1982reflection}. For example, the $C_R$ value in  Eq. \eqref{eq:Davies82} is always less than unity, 
which limits the effect of $m$. Hence the patch length after a certain threshold has practically no effect, exactly as observed in Figs.\, \ref{fig:patch_size_diff}(a)--\ref{fig:patch_size_diff}(b) for $2l\geq \pi/2$. In summary, increasing the patch length increases negative drift (especially effective for particles in the vicinity of the first bottom ripple), however the effect saturates after a threshold patch length is reached.


\section{Modified Stokes drift due to the combined effect of  bottom ripples and  background current}
\label{sec:with_current}

Up to this point, we have not considered the effect of background current, which in oceanic scenarios inevitably affects the Lagrangian drift. 
In the presence of a background current, bottom ripples introduce a time-independent non-homogeneous solution  to the system.
The total perturbed velocity, i.e.\, homogeneous solution (denoted by subscript `$h$') $+$  particular integral (denoted by subscript `$p$'),  
at $\mathcal{O}(\epsilon)$ for a pair of counter-propagating surface waves (with respective amplitudes $a$ and $Ra$ and respective frequencies $\omega$ and $-\omega$) in the presence of constant  background current $U_0$ and bottom ripples of the form $\eta_b(x)=a_b \sin(k_b x)$ can be written as: 
\begin{subequations}
\begin{align}
u^{(1)}&= u_h^{(1)}+u_p^{(1)},    \\
w^{(1)}&= w_h^{(1)}+w_p^{(1)},
\end{align}
\end{subequations}
where
\begin{align*}
    u_h^{(1)} & =a \overline{\omega} \frac{\cosh k(z+H)}{\sinh(kH)}\cos(kx-\omega t) -a R \overline{\omega} \frac{\cosh k(z+H)}{\sinh(kH)}\cos(kx+\omega t),\\
    u_p^{(1)}& =-a_b k_b \Big[\mathcal{C}_1 \frac{\cosh k_b(z+H)}{\cosh(k_b H)}  +  \mathcal{C}_2\frac{\sinh(k_b z)}{\cosh(k_b H)} \Big] \cos(k_b x),\\
    w_h^{(1)}&=a \overline{\omega} \frac{\sinh k(z+H)}{\sinh(kH)}\sin(kx-\omega t)-a R \overline{\omega} \frac{\sinh k(z+H)}{\sinh(kH)}\sin(kx+\omega t),\\
    w_p^{(1)}&=-a_b k_b \Big[\mathcal{C}_1 \frac{\sinh k_b(z+H)}{\cosh(k_b H)}  + \mathcal{C}_2\frac{\cosh(k_b z)}{\cosh(k_b H)} \Big] \sin(k_b x).
\end{align*}
Here $\mathcal{C}_1=- U_0 k_b^2 g[\det(\mathcal{D})(0,k_b) \cosh(k_b H)]^{-1}$, $\mathcal{C}_2 =  U_0$,  $\overline{\omega}=\omega-U_0 k$, and $\det(\mathcal{D})(0,k_b)=k_b[g k_b \tanh(k_b H) - U_0^2 k_b^2]$.  Details about the derivation of the velocity potential, from which the velocity field has been obtained, are given in Appendix \ref{Sec:amp_evol} (see specifically Eq. \eqref{app eq:phi_1}). We mention in passing that surface wave and rippled bottom topography interaction in the presence of a constant current has been elaborately studied in \citet{kirby1988current} in the context of Bragg resonance.

We proceed following the steps outlined  in Sec. \ref{counter_waves} and calculate $\boldsymbol{\xi}^{(1)}$:
\begin{subequations}
\begin{align}
\xi^{(1)}&= \xi_h^{(1)}+\xi_p^{(1)} ,   \\
\zeta^{(1)}&= \zeta_h^{(1)}+\zeta_p^{(1)},
\end{align}
\end{subequations}
where
\begin{align*}
\xi_h^{(1)} & = - \frac{a \overline{\omega}}{\omega} \frac{\cosh k(z_o+H)}{\sinh(kH)}\sin(kx_o-\omega t)- \frac{a R \overline{\omega}}{\omega} \frac{\cosh k(z_o+H)}{\sinh(kH)}\sin(kx_o+\omega t),\\
\xi_p^{(1)}&=- t a_b k_b \Big[\mathcal{C}_1 \frac{\cosh k_b(z_o+H)}{\cosh(k_b H)}  + \mathcal{C}_2\frac{\sinh(k_b z_o)}{\cosh(k_b H)} \Big] \cos(k_b x_o),\\
\zeta_h^{(1)}&=\frac{a \overline{\omega}}{\omega} \frac{\sinh k(z_o+H)}{\sinh(kh)}\cos(kx_o-\omega t)+\frac{a R \overline{\omega}}{\omega} \frac{\sinh k(z_o+H)}{\sinh(kh)}\cos(kx_o+\omega t),\\
\zeta_p^{(1)}&= -  t a_b k_b \Big[\mathcal{C}_1 \frac{\sinh k_b(z_o+h)}{\cosh(k_b H)}  + \mathcal{C}_2\frac{\cosh(k_b z_o)}{\cosh(k_b H)} \Big] \sin(k_b x_o) .
\end{align*}

For calculating the \emph{modified} Stokes drift, we proceed as in Sec. \ref{subsec:stokes_drift}:
\begin{equation}
   \begin{bmatrix}
u_{SD} &  w_{SD} 
\end{bmatrix}
=
\begin{bmatrix}
\xi_h^{(1)}+\xi_p^{(1)} &  \zeta_h^{(1)}+\zeta_p^{(1)} 
\end{bmatrix}
\left. \begin{bmatrix}
u_{h,x}^{(1)} +u_{p,x}^{(1)} & w_{h,x}^{(1)}+w_{p,x}^{(1)}\\ \\ 
u_{h,z}^{(1)} +u_{p,z}^{(1)} & w_{h,z}^{(1)}+w_{p,z}^{(1)}
\end{bmatrix}  \right \vert_{(x_0,z_0)}.
\label{eq:Stokes_drift_formula1}
\end{equation}
Time averaging over one time-period finally leads to the \emph{modified} Stokes drift $\langle \boldsymbol{u_{SD}}\rangle$:
\begin{subequations}
\begin{align}
\langle u_{SD} \rangle &= \underbrace{\langle\xi_h^{(1)} u_{h,x}^{(1)}+ \zeta_h^{(1)} u_{h,z}^{(1)}\rangle}_{\text{\clap{Term-1}}}+  \underbrace{ \langle\xi_h^{(1)} u_{p,x}^{(1)}+ \zeta_h^{(1)} u_{p,z}^{(1)}\rangle}_{\text{\clap{Term-2}}} + \nonumber \\& \underbrace{\langle \xi_p^{(1)} u_{h,x}^{(1)}+ \zeta_p^{(1)} u_{h,z}^{(1)} \rangle}_{\text{\clap{Term-3}}}+ \underbrace{\langle\xi_p^{(1)} u_{p,x}^{(1)}+ \zeta_p^{(1)} u_{p,z}^{(1)}\rangle}_{\text{\clap{Term-4}}}, \label{app eq:u_sd}\\
\langle {w_{SD}}\rangle &=\underbrace{\langle \xi_h^{(1)} w_{h,x}^{(1)}+ \zeta_h^{(1)} w_{h,z}^{(1)}\rangle}_{\text{\clap{Term-5}}}+\underbrace{\langle \xi_h^{(1)} w_{p,x}^{(1)}+\zeta_h^{(1)} w_{p,z}^{(1)}\rangle}_{\text{\clap{Term-6}}}+ \nonumber \\&\underbrace{\langle\xi_p^{(1)} w_{h,x}^{(1)}+ \zeta_p^{(1)} w_{h,z}^{(1)}\rangle}_{\text{\clap{Term-7}}}+\underbrace{\langle\xi_p^{(1)} w_{p,x}^{(1)}+ \zeta_p^{(1)} w_{p,z}^{(1)}\rangle}_{\text{\clap{Term-8}}}, \label{app eq:w_sd}
\end{align}
\end{subequations}
where
\begin{align*}
\text{Term-1}&= \frac{a^2(1-R^2) \overline{\omega}^2 k}{2 \omega \sinh^2{(kH)}}\cosh[2k(z_o+H)], \\
\text{Term-2}&=0,\\
\text{Term-3}&=\frac{a_b U_0 k_b a (1+R) k}{\omega} \overline{\omega} \Bigg[ \mathcal{X} \cos(k x_o)\cos(k_b x_o) \frac{\cosh k(z_o+H)}{\sinh(k H)}  + \nonumber \\ & \hspace{6cm}\mathcal{Z} \sin(k x_o) \sin(k_b x_o) \frac{\sinh k(z_o+H)}{\sinh(k H)} \Bigg],\\
\text{Term-4}&=-\frac{\pi a_b^2 U_0^2 k_b^3 }{ \omega} (\mathcal{X}^2-\mathcal{Z}^2) \sin(k_b x_o)\cos(k_b x_o),
\end{align*}

\begin{align*}
\text{Term-5}&=0,\\
\text{Term-6}&=0,\\
\text{Term-7}&=\frac{a_b U_0 k_b a(1+R) k}{\omega} \overline{\omega} \Bigg[ \mathcal{X} \sin(k x_o)\cos(k_b x_o) \frac{\sinh k(z_o+H)}{\sinh(k H)} - \hspace{2cm} & \nonumber \\&   \hspace{6cm} \mathcal{Z} \cos(k x_o) \sin(k_b x_o) \frac{\cosh k(z_o+H)}{\sinh(k H)} \Bigg],\\
\text{Term-8}&=\frac{\pi a_b^2 U_0^2 k_b^3 }{\omega} \mathcal{X} \mathcal{Z},
\end{align*}
in which
\begin{align*}
    \mathcal{X}&=\frac{1}{U_0}\Big[\mathcal{C}_1 \dfrac{\cosh k_b(z_o+H)}{\cosh(k_b H)}  + \mathcal{C}_2\dfrac{\sinh(k_b z_o)}{\cosh(k_b H)} \Big],\\
    \mathcal{Z}&=\frac{1}{U_0}\Big[\mathcal{C}_1 \dfrac{\sinh k_b(z_o+H)}{\cosh(k_b H)}  + \mathcal{C}_2\dfrac{\cosh(k_b z_o)}{\cosh(k_b H)} \Big].
\end{align*}

Comparison between Eqs. \eqref{app eq:u_sd}--\eqref{app eq:w_sd}  and  Eqs. \eqref{z_SDa}--\eqref{z_SDb}  reveals that the  Stokes drift for two counter-propagating waves (or for a single wave, i.e. when $R=0$) is substantially more involved in the presence of a constant mean current and bottom ripples. Term-1 in Eq. \eqref{app eq:u_sd} directly corresponds to Eq. \eqref{z_SDa} while  Term-5 in Eq. \eqref{app eq:w_sd} (which is zero) is exactly the same as Eq. \eqref{z_SDb}. This is evident since Term-1 and Term-5 are purely homogeneous solutions. All other terms in Eqs. \eqref{app eq:u_sd}--\eqref{app eq:w_sd} involve particular solutions arising from the interactions between the mean current and the bottom ripples.

The order of magnitude of each of the terms arising in Eqs. \eqref{app eq:u_sd}--\eqref{app eq:w_sd} can be straight-forwardly compared for the long--wave, long--ripple limit,  {i.e. when $\alpha \equiv kH \ll \mathcal{O}(1)$ and $\alpha_b \equiv k_bH \ll \mathcal{O}(1)$.} Non-dimensionalizing $\langle u_{SD} \rangle$ and $\langle w_{SD} \rangle$ with long wave speed $\sqrt{gH}$ yields for Froude number, $Fr \equiv U_0/\sqrt{gH}=\mathcal{O}(1)$:
\begin{eqnarray}
\text{Term-1}=\mathcal{O}(\Tilde{a}^2),\,\text{Term-3}=\mathcal{O}(\Tilde{a}\Tilde{a}_b),\,\text{Term-4}=\mathcal{O}( \Tilde{a}_b^2 \alpha_b\alpha^{-1}), \nonumber\\
\text{Term-7}=
\mathcal{O}(\Tilde{a}\Tilde{a}_b \alpha) + \mathcal{O}(\Tilde{a}\Tilde{a}_b \alpha_b)\,\,\,\mathrm{and}\,\,\text{Term-8}=\mathcal{O}( \Tilde{a}_b^2 \alpha_b^2\alpha^{-1}),
\label{eq:OMAnalysis}
\end{eqnarray}
where each of the `Term-n's has been non-dimensionalized by $\sqrt{gH}$,
$\Tilde{a}\equiv a/H$ and $\Tilde{a}_b\equiv a_b/H$.
The assumptions $Fr=\mathcal{O}(1)$,  $\alpha \ll \mathcal{O}(1)$ and $\alpha_b \ll \mathcal{O}(1)$ are relevant in the coastal oceanic context.
The comparative orders of magnitude of the two components of Term-7  are not known \emph{a-priori}. While the steepness parameter must yield $\mathcal{O}(\alpha\Tilde{a})=\mathcal{O}(\alpha_b\Tilde{a}_b)$ (see Appendix \ref{Sec:amp_evol}), there is no implicit assumption that  $\mathcal{O}(\alpha)=\mathcal{O}(\alpha_b)$ and $\mathcal{O}(\Tilde{a})=\mathcal{O}(\Tilde{a}_b)$. However the assumption $\mathcal{O}(\alpha)=\mathcal{O}(\alpha_b)$ (which then yields $\mathcal{O}(\Tilde{a})=\mathcal{O}(\Tilde{a}_b)$) is particularly helpful for comparing the order of magnitude of  different terms.
 Under these assumptions we have:
\begin{equation*}
    \text{Term-1} \sim \text{Term-3} \sim \text{Term-4} \gg \text{Term-7} \sim \text{Term-8}.
\end{equation*}
We recall here that Term-1, -3 and -4 arise from the horizontal drift, and hence  are expected to be much greater than Term-7 and -8, which arise from the vertical drift. Hence the straight-forward order of magnitude analysis has revealed that for $Fr=\mathcal{O}(1)$ and $\mathcal{O}(\alpha)=\mathcal{O}(\alpha_b)\ll \mathcal{O}(1)$, Term-3 and -4 arising from the particular solution are comparable with the homogeneous solution, Term-1 (which for $R=0$ leads to the classical Stokes drift).

Next we try to understand the different terms arising from the particular solution.
Term-2 and Term-6, which are both zero,  are the time averages of the inner products of (i) the particle displacement due to surface waves (which is periodic in time) and (ii) the time-independent velocity gradient  arising from the interactions between the constant current and bottom ripples.   Term-3 and Term-7 are the time averages of the inner products of (i) the particle displacement due to the interactions between the constant current and bottom ripples (which is linear in time) and (ii) the  velocity gradient  arising from the surface waves (which is periodic in time). Finally Term-4 and Term-8 are the time averages of the inner products of (i) the particle displacement due to the interactions between the constant current and bottom ripples (which is linear in time) and (ii) the corresponding velocity gradient (which is time independent). We emphasize the obvious fact that the effect of the particular integrals is independent of  whether the bottom ripples satisfy the Bragg resonance condition. In other words, these additional terms would always arise when shallow or intermediate depth surface waves propagate over corrugated sea-floor in the presence of a mean current.

In the presence of reflected waves, 
we can symbolically express the modified Stokes drift as follows:  
\begin{equation*}
    \mathrm{Modified\,Stokes\, drift}=\underbrace{\mathrm{Stokes\, drift_{incident}}+\mathrm{Stokes \,drift_{reflected}}}_{\text{\clap{including particular integral}}}.
\end{equation*}

\begin{figure}
    \centering
    \includegraphics[width=10cm]{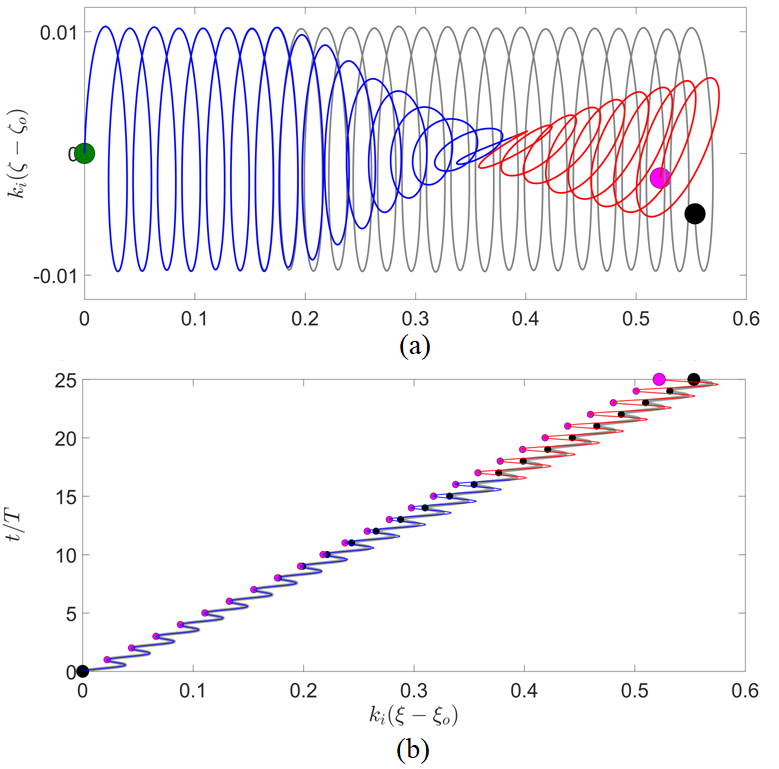}
    \caption{Comparison of trajectories of a parcel initially located at \textcircled{b} in the presence of a constant background current $Fr=0.003$  with and without bottom ripples.
    (a) Lagrangian trajectory without (gray) and with (blue and red) bottom ripples. Blue  trajectories denote forward (clockwise) drift, while red trajectories shows backward (anti-clockwise) drift. Green circle denotes the initial position of the parcel, while black and magenta  circles respectively denote final positions without and with bottom ripples.  
    (b) Horizontal component of Lagrangian-mean trajectories without  (black dotted) and with (magenta dotted) bottom ripples.  Each dot represents position after one time-period.}
    \label{fig:particle_traj_with_U}
\end{figure}

Next we consider an `unfavourable situation', i.e. $U_0>0$ and is therefore aligned in the direction of the incident wave, thus fully helping the forward drift.  To compare with realistic oceanic scenarios, we choose nearshore depth of $H=50$m and $U_0\approx 0.07$m/s (from surface currents data in \citet{dohan2017ocean}) such that $Fr=0.003$. The Bragg resonance condition is nearly satisfied with $k_b=2k_i$ (small deviations occur due to the weak mean current), and intermediate depth condition is accomplished by assuming $k_iH=1$. The bottom ripple follows Eq. \eqref{eq:baines_profile} with $2l=\pi/2$ and $a_b=0.1H$. Note that the magnitudes of various parameters chosen here are different from the order of magnitude analysis in Eq. \eqref{eq:OMAnalysis}, which is valid for the long wave, long ripple, and $Fr=\mathcal{O}(1)$ limit (which allowed a simple order of magnitude comparison that is not apparent for intermediate depths).

Bragg resonance  leads to the generation of reflected waves; the  amplitude evolution equations for incident and reflected waves (for the simplified case of patch with infinite horizontal extent) are given by Eq. \eqref{eq:final_amp_evol}.
Figure \ref{fig:particle_traj_with_U}(a) shows a comparison between Lagrangian trajectories of particle \textcircled{b} in the presence and absence of  bottom ripples, 
while Fig.\, \ref{fig:particle_traj_with_U}(b) compares the $x$-component of the Lagrangian-mean trajectories.
Noting that 
\begin{equation*}
    \mathrm{Lagrangian}=\mathrm{Eulerian}(U_0)+\mathrm{Modified\,Stokes\, drift},
\end{equation*}
 the forward Lagrangian-mean motion  is  hindered by the reflected waves generated by the bottom ripple, as is evident from the difference between the dotted black and magenta lines in Fig.\, \ref{fig:particle_traj_with_U}(b).
The backward drift generated by the bottom ripples, while not sufficient to cause particle trapping or reflection, is enough to create $\approx 6\%$ difference within the first $25$ time-periods. The trends  reveal that the difference between the two Lagrangian-mean trajectories almost linearly grows in time. The backward drift due to Bragg reflection is expected to be larger for larger values of $a_b$ (see Fig.\, \ref{fig:different_ab}).

We emphasize here that careful analysis in \citet{kirby1988current} revealed that  the presence of a  mean current of arbitrary sign uniformly  enhances resonant reflection of  incident waves due to the additional effects arising from the current-topography interactions. In our modified Stokes drift context, this implies that the backward drift would always be enhanced independent of the sign of the mean current.


\section{Summary and Conclusion}
\label{summary}
The seminal work of Stokes \cite{stokes1847} showed that surface waves cause mass transport; particles undergo a slow drift in the direction of propagation of the waves. This theory is central to understanding how material parcels, e.g.\, pollutants, sediments, plankton, etc., are transported in aquatic environments. There exists a vast literature on Stokes drift, especially in the deep-water limit.
However to the best of our knowledge, the effect of sea-floor corrugations on the drift velocity (in the presence or absence of mean current),  particularly relevant for intermediate and shallow depths, has not been explored previously. 
This may be crucially important for understanding how pollutants/effluents from river discharges make its way into deeper oceans by  travelling through the shallow nearshore regions. Therefore, the objective of this paper is  to fundamentally understand wave-induced tracer transport in coastal oceans, and in this regard we have investigated  how Stokes drift is affected when intermediate depth surface waves  resonantly interact with bottom ripples, both in the presence and absence of a constant mean current.

We have first considered the simplest setting of two counter-propagating Stokes waves (with amplitudes $a$ and $Ra$) in a fluid of constant depth, and we have analytically demonstrated that Stokes drift diminishes as $R$ increases (for $0\le R \le1$) and becomes zero when $R=1$. Furthermore, for $R>1$, the drift again increases but in the negative direction. Counter-propagating waves may arise when a surface wave travels over a corrugated bottom topography; nonlinear resonant interaction between an incident surface wave and a rippled bottom gives rise to a reflected wave. The theoretical underpinnings of this process -- the Bragg resonance -- has been described next; an incident wave $(k_i,\omega_i)$, mediated by a bottom ripple of wavenumber $k_b=2k_i$,  transfers energy (and vice versa) to a reflected wave $(k_r,\omega_r)=(k_i,-\omega_i)$. The energy transfer is more pronounced in the shallow, nearshore region of the oceans.  Next, using the HOS method, we have numerically investigated how  a finite patch of bottom ripples affects  Stokes drift. When the incident wave is rightward propagating and the bottom is flat, we expect a unidirectional Stokes drift (rightward in this case) with clockwise circulating particle trajectories.  The rippled patch, however, leads to Bragg reflection, rendering the free surface a complex interference of incident and reflected waves.  When the induced velocity due to the reflected waves is stronger than the incident, the drift is backward (leftward), and the particle trajectories have an anti-clockwise circulation, which is analogous to $R>1$ case in our model theoretical analysis with Stokes waves.
We have demonstrated the reversal of Stokes drift by studying Lagrangian and Lagrangian-mean trajectories of multiple particles placed on the free surface above the patch. Hence we conclude that the rippled patch acts as a non-surface-invasive particle trap or reflector.

The number of bottom ripples (or patch length) and its amplitude have a significant effect on particle trajectories — the amplitude of the reflected wave increases as the ripple amplitude or patch length increases.
For the latter, however, the amplitude of the reflected waves do not increase indefinitely and saturates after a threshold patch length is reached. This may provide a design estimate of the effective length of an artificial rippled patch. Lagrangian-mean trajectories reveal that particles in the vicinity of the up-wave side of the rippled-patch are either trapped, slowed down considerably, or reflected.

Finally we considered the effect of a constant mean current over a rippled bottom topography on Stokes drift. We derived  the modified Stokes drift for two counter-propagating surface waves in a constant mean current over a rippled bottom topography of infinite horizontal extent. New terms  arising from the particular integral (due to the interactions between the constant mean current and rippled bottom topography) appear in this
\emph{modified} Stokes drift, which reveals how Stokes drift in this scenario is fundamentally more complex from the one that is commonly known. 
Since the effect of the particular integrals is independent of  whether the bottom ripples satisfy the Bragg resonance condition, these additional terms would always arise when shallow or intermediate depth surface waves propagate over corrugated sea-floor in the presence of a mean current.

We  numerically  investigated the 
`unfavourable situation' of a constant mean current of realistic magnitude aligned in the direction of the incident wave, thereby favoring the forward drift. With  ripple amplitude   $10\%$ of the water depth, the backward drift due to  Bragg reflection, in spite of the combined effects of forward mean advection and forward drift due to the incident wave, is able to create $\approx 6\%$ difference between the horizontal component of the Lagrangian-mean trajectories (i.e. difference observed between the presence and absence of the rippled patch) within the first $25$ time-periods. In fact, this difference almost linearly grows with time, further endorsing the key role of corrugated bathymetry.

While we limited our study to a monochromatic incident wave, in real ocean scenarios, an incident wave-group contains a spectrum of frequencies and is composed of many (linearly or nonlinearly) superposed wave components. Motivated by wave energy harvesting and coastal protection applications, \citet{elandt2014surface} studied Bragg resonance (specifically, surface wave lensing) using a broadband spectrum of incident waves and polychromatic topography.  Complexity does arise because of super/sub-harmonic generations, quartet, and higher-order Bragg resonances, however, the underlying mechanism of lensing of monochromatic and broadband waves is similar. Hence we expect that in realistic scenarios, a properly designed artificial patch of bottom ripple can serve multiple purposes -- from wave energy harvesting and coastal protection to pollutant trapping.  One application of our work could be mitigating microplastic pollution, which is a growing concern and a severe threat to the marine environment. Studies suggest that mechanical degradation of plastics occurs in coastal areas, leading to microplastics, which then makes its onward journey into the deep sea \cite{chubarenko2018behavior}. Topographic interactions are inevitable in the nearshore region, and our study has revealed that sea-floor corrugations, which might seem unimportant, can significantly affect  cross-shelf exchange of microplastics and other nearshore tracers like pathogens, contaminants, nutrients, larvae and sediments. Hence high-resolution coastal bathymetry map needs to be included for predicting pollutant pathways.




\begin{acknowledgments}
The authors thank R. Raj for useful discussions and G. Saranraj for carefully reading the manuscript, and providing useful comments and suggestions.
\end{acknowledgments}


\appendix

\section{Amplitude evolution equations of surface waves with constant current and rippled bottom bathymetry}
\label{Sec:amp_evol}

For completeness, we follow \citet{kirby1988current} to derive (with a  simplified consideration and slightly different approach) the amplitude evolution equations for Bragg resonance in the presence of a constant current. 

First we perform a perturbation expansion of $\eta$ and $\phi$:
\renewcommand{\theequation}{A \arabic{equation}}
\begin{subequations}
\begin{align}
& \eta= \eta^{(1)}+ \eta^{(2)}+\mathcal{O}(\epsilon^3), \label{eq:eta_exp}\\
& \phi= \phi^{(1)}+ \phi^{(2)}+\mathcal{O}(\epsilon^3), \label{eq:phi_exp}
\end{align}
\end{subequations}
where superscript `$(n)$' denotes $\mathcal{O}(\epsilon^n)$, $\epsilon$ being a small parameter scaling with the wave steepness.
We collect all the terms corresponding to $\mathcal{O}(\epsilon)$ appearing in the governing Laplace equation Eq. \eqref{eq:Gov_eq} and the  boundary conditions Eq. \eqref{eq:kin1_modified}, Eqs. \eqref{eq:kin2}--\eqref{eq:dyn}:
\begin{subequations}
\begin{align}
&[\mathrm{GE}]: \hspace{1.1cm} \nabla^2\phi^{(1)}=0,\\
&[\mathrm{KBC}]: \hspace{0.9cm} \eta^{(1)}_{,t}+ U_0\eta^{(1)}_{,x}-\phi^{(1)}_{,z}=0   \qquad &\mathrm{at} \, z=0,\\
&[\mathrm{ImC}]:  \hspace{1cm} \phi^{(1)}_{,z}=\eta_{b_{,x}}U_0   \qquad &\mathrm{at} \, z=-H, \label{App eq:IMC_1} \\
&[\mathrm{DBC}]:  \hspace{0.9cm} \phi^{(1)}_{,t}+U_0 \phi^{(1)}_{,x}+ g\eta^{(1)}=0 \qquad &\mathrm{at} \, z=0.
\end{align}
\end{subequations}
Impenetrability condition (ImC) in presence of background current leads to an inhomogeneous boundary condition. We define the solution of the surface elevation and the velocity potential as a combination of homogeneous solution and particular integral:
\begin{eqnarray}
  \eta^{(1)}=\eta^{(1)}_h+\eta^{(1)}_p, \\
  \phi^{(1)}=\phi^{(1)}_h+\phi^{(1)}_p.
\end{eqnarray}
The homogeneous solution for two modes, $k_1$ and $k_2$, are given as follows:
\begin{subequations}
\begin{align}
\eta_h^{(1)}(x,t,\tau)&=\sum_{j=1}^{2} a_j(\tau) \ee^{\ii(k_j x-\omega_j t)}+\textnormal{c.c.}, \label{app eq:eta_h} \\
\phi_h^{(1)}(x,t,\tau)&=\sum_{j=1}^{2} \Big[A_j \dfrac{\cosh k_j(z+H)}{\cosh(k_j H)}  + B_j\dfrac{\sinh(k_j z)}{\cosh(k_j H)} \Big] \ee^{\ii(k_j x-\omega_j t)}+\textnormal{c.c.}, \label{app eq:phi_h}
\end{align}
\end{subequations}

where c.c. denotes the complex conjugate and  $\tau=\epsilon t$ is a slow time scale. The bottom topography profile reads
\begin{equation}
    \eta_b(x)=a_b \, \ee^{\ii k_b x}+\textnormal{c.c.} \label{app eq:etab}
\end{equation}
An assumption implicit in Eq. \eqref{App eq:IMC_1} is that for $\mathcal{O}(1)$ mean current, the steepness of the bottom ripple, $k_ba_b$ should be of the same order of magnitude as $\epsilon$.

Using  Eqs. \eqref{app eq:eta_h}--\eqref{app eq:phi_h}, we finally obtain a system of equations:

\begin{equation}
\mathcal{D}(\omega_j,k_j) \mathbf{x}_j^{(1)}=0 , \hspace{10mm}  j=1,2 \label{eq:matrix}
\end{equation}
where,\\
$\mathcal{D}(\omega_j,k_j)= $
$\begin{bmatrix}
k_j \tanh(k_j H) & \dfrac{k_j}{\cosh(k_j H)} & \ii \overline{\omega}_j\\
0 & k_j & 0 \\
-\ii \overline{\omega}_j & 0 & g
\end{bmatrix}$, 
$\mathbf{x}_j^{(1)}=$
$\begin{bmatrix}
A_j,
B_j,
a_j
\end{bmatrix}^\dagger$ and $\overline{\omega}_j\equiv \omega_j-U_0 k_j$.

The homogeneous solution for the velocity potential is readily found to be:
\begin{eqnarray}
\phi_h^{(1)}(x,t,\tau)= \sum_{j=1}^{2} \Big[\frac{-\ii g a_j }{\overline{\omega}_j} \frac{\cosh k_j(z+H)}{\cosh(k_j H)}\Big] \ee^{\ii(k_j x-\omega_j t)}.
\end{eqnarray}
We also calculate the null vector of  the matrix $\mathcal{D}(\omega_j,k_j)^\dagger$:
\begin{equation}
\mathbf{n_j}=
\begin{bmatrix}
1,
\dfrac{1}{\cosh(k_j H)},
\dfrac{\ii \overline{\omega}_j}{g}
\end{bmatrix}^\dagger.   
\label{eq:nj}
\end{equation}

Non-homogeneous bottom boundary condition (ImC) introduces a time-independent particular solution, which can be calculated by  assuming the particular solution of the surface elevation and velocity potential as follows:
\begin{subequations}
\begin{align}
\eta_p^{(1)}(x)&=\hat{a} \ee^{\ii k_b x}+\textnormal{c.c.}, \label{app eq:eta_p}\\
\phi_p^{(1)}(x)&=\Big[\hat{A} \frac{\cosh k_b(z+H)}{\cosh(k_b H)}+ \hat{B} \frac{\sinh(k_b z)}{\cosh(k_b H)}\Big] \ee^{\ii k_b x}+\textnormal{c.c.} \label{app eq:phi_p}
\end{align}
\end{subequations}

Substitution of  Eqs. \eqref{app eq:eta_p}--\eqref{app eq:phi_p} in the boundary conditions yield:
\begin{subequations}
\begin{align}
&[\mathrm{KBC}]: \hspace{0.9cm} \Big[\ii U_0 \hat{a} k_b -k_b \Big\{\hat{A} \tanh(k_b H)-\dfrac{\hat{B}}{\cosh(k_b H)}\Big\} \Big] \ee^{\ii k_b x}=0,\\
&[\mathrm{ImC}]:  \hspace{1.1cm} k_b \hat{B} = \ii a_b k_b U_0,\\
&[\mathrm{DBC}]:  \hspace{1cm} [\ii k_b U_0 \hat{A}+g \hat{a} ]  \ee^{\ii k_b x}=0,
\end{align}
\end{subequations}

which can be concisely written as the following matrix problem:
\begin{equation}
\mathcal{D}(0,k_b) \mathbf{\hat{x}}^{(1)}=\mathbf{\hat{b}}, 
\end{equation}
where $\mathcal{D}(\omega_b=0,k_b)$ is obtained on substituting  $\omega_j=0$ and $k_j=k_b$ in Eq. \eqref{eq:matrix}. Furthermore,
$\mathbf{\hat{x}}^{(1)}=$
$\begin{bmatrix}
\hat{A},
\hat{B},
\hat{a}
\end{bmatrix}^\dagger$ and
$\mathbf{\hat{b}} =$
$\begin{bmatrix}
0,
\ii a_b k_b U_0,
0
\end{bmatrix}^\dagger$.

The total solution of the surface elevation and velocity potential at $\mathcal{O}(\epsilon)$ in the presence of the bottom topography and background current are therefore as follows:
\begin{subequations}
\begin{align}
\eta^{(1)}(x,t,\tau)=&\sum_{j=1}^{2} a_j(\tau) \ee^{\ii(k_j x-\omega_j t)}+\mathcal{A}_b a_b \ee^{\ii k_b x}+\textnormal{c.c.},\label{app eq:eta_1}\\
\phi^{(1)}(x,t,\tau)=& \sum_{j=1}^{2} \Big[\frac{-\ii a_j g}{\overline{\omega}_j} \frac{\cosh k_j(z+H)}{\cosh(k_jH)}\Big] \ee^{\ii(k_j x-\omega_j t)}+\nonumber \\ & \hspace{10mm} a_b \Big[\mathcal{A} \frac{\cosh k_b(z+H)}{\cosh(k_b H)}  + \mathcal{B} \frac{\sinh(k_b z)}{\cosh(k_b H)} \Big] \ee^{\ii k_b x}+\textnormal{c.c.}, \label{app eq:phi_1}
\end{align}
\end{subequations}
where \begin{align*}
   \mathcal{A} & =-\dfrac{\ii U_0 k_b^2 g }{\det(\mathcal{D})(0,k_b) \cosh(k_b H)},\\
    \mathcal{B} & =\ii  U_0,\\
    \mathcal{A}_b & =-\dfrac{ U_0^2 k_b^3}{ \det(\mathcal{D})(0,k_b) \cosh(k_b H)},
\end{align*}
and $\det(\mathcal{D})(0,k_b)=k_b[g k_b \tanh(k_b H) - U_0^2 k_b^2]$ is the determinant of the matrix $\mathcal{D}$.

Similarly, we can collect all the terms corresponding to $\mathcal{O}(\epsilon^2)$:
\begin{subequations}
\begin{align}
&[\mathrm{GE}]: \hspace{6mm} \nabla^2\phi^{(2)}=0 \label{eq:laplace_2},\\
&[\mathrm{KBC}]: \hspace{4mm} \eta^{(2)}_{,t}+ U_0\eta^{(2)}_{,x}-\phi^{(2)}_{,z}=- \eta^{(1)}_{,x} \phi^{(1)}_{,x}+\eta^{(1)}\phi^{(1)}_{,zz}-\eta^{(1)}_{,\tau}, \qquad  \mathrm{at} \, z=0, \label{eq:KBC_2} \\
&[\mathrm{ImC}]: \hspace{5mm} \phi^{(2)}_{,z} = \eta_{b_x} \phi^{(1)}_{,x}-\eta_{b} \phi^{(1)}_{,zz} ,    \hspace{54mm} \mathrm{at} \, z=-H, \label{eq:ImC_2}\\
&[\mathrm{DBC}]: \hspace{4mm} \phi^{(2)}_{,t}+ U_0 \phi^{(2)}_{,x}+ g\eta^{(2)}=-\frac{1}{2}[(\phi^{(1)}_{,x})^2+(\phi^{(1)}_{,z})^2]-\eta^{(1)}(\phi^{(1)}_{,tz}+U_0  \phi^{(1)}_{,xz})-\phi^{(1)}_{,\tau}, \qquad \nonumber \\& \hspace{114mm} \mathrm{at} \, z=0. \label{eq:DBC_2}
\end{align}
\end{subequations}
The homogeneous solution at $\mathcal{O}(\epsilon^2)$ for modes $k_1$ and $k_2$ are given as follows: 
\begin{subequations}
\begin{align}
\eta^{(2)}(x,t,\tau)&=\sum_{j=1}^{2} d_j(\tau) \ee^{2\ii(k_j x-\omega_j t)}+\textnormal{c.c.}, \label{app eq:eta_2}\\ 
\phi^{(2)}(x,t,\tau)&=\sum_{j=1}^{2} \Big[C_j(\tau) \frac{\cosh 2k_j(z+H)}{\cosh(2k_jH)}+ D_j(\tau) \frac{\sinh(2k_j z)}{\cosh(2k_jH)}\Big] \ee^{2\ii(k_j x-\omega_j t)}+\textnormal{c.c.}
\label{app eq:phi_2}
\end{align}
\end{subequations}

Using  Eqs. \eqref{app eq:eta_2}--\eqref{app eq:phi_2} and the resonance conditions $k_1+k_2=k_b$ and $\omega_1+\omega_2=0$, the following system of equations are obtained:
\begin{subequations}
\begin{align}
\mathcal{D}(\omega_1,k_1) \mathbf{x_1}^{(2)}=\mathbf{v_1}a_b \overline{a}_2^{(1)}+ \mathbf{r_1} a^{(1)}_{1,\tau},\\
\mathcal{D}(\omega_2,k_2) \mathbf{x_2}^{(2)}=\mathbf{v_2}a_b \overline{a}_1^{(1)}+ \mathbf{r_2} a^{(1)}_{2,\tau},
\end{align}
\end{subequations}
where
\begin{align*}
\mathbf{x}_j^{(2)}&=
\begin{bmatrix}
C_j,
D_j,
d_j
\end{bmatrix}^\dagger,\\
\mathbf{v_1}&=
\begin{bmatrix}
k_1 \Big[\mathcal{A} k_b+ \dfrac{\ii k_2 \mathcal{A}_b g}{\overline{\omega}_2}\Big]\\
\dfrac{-\ii k_1 k_2 g}{\overline{\omega}_2 \cosh(k_2 H)}\\
\dfrac{\ii g k_2 k_b \mathcal{A}}{\overline{\omega}_2}-\overline{\omega}_2^2 \mathcal{A}_b+ \ii k_b (\overline{\omega}_2-U_0 k_b) \Big[\dfrac{\mathcal{B}}{\cosh(k_b H)}+ \mathcal{A} \tanh(k_b H)\Big]
\end{bmatrix},\\
\mathbf{v_2}&=
\begin{bmatrix}
k_2 \Big[\mathcal{A} k_b+ \dfrac{\ii k_1 \mathcal{A}_b g}{\overline{\omega}_1}\Big]\\
\dfrac{-\ii k_1 k_2 g}{\overline{\omega}_1 \cosh(k_1 H)}\\
\dfrac{\ii g k_1 k_b \mathcal{A}}{\overline{\omega}_1}-\overline{\omega}_1^2 \mathcal{A}_b+ \ii k_b (\overline{\omega}_1-U_0 k_b) \Big[\dfrac{\mathcal{B}}{\cosh(k_b H)}+ \mathcal{A} \tanh(k_b H)\Big]
\end{bmatrix},\\
\mathbf{r_j}&=
\begin{bmatrix}
-1,
0,
\dfrac{\ii g}{\overline{\omega}_j}
\end{bmatrix}^\dagger.
\end{align*}

Using the Fredholm alternative, we obtain the amplitude evolution equations:
\begin{equation}
\frac{da_1}{d\tau}= \beta_1 a_b\bar{a}_2\qquad; \hspace{5mm} \frac{da_2}{d\tau}= \beta_2 a_b\bar{a}_1,
\label{eq:final_amp_evol}
\end{equation}
where
\begin{equation*}
\beta_j = -\frac{\mathbf{n_j \cdot v_j}}{\mathbf{n_j \cdot r_j}},
\end{equation*}
in which $\mathbf{n_j}$ is obtained from Eq. \eqref{eq:nj}. The superscript `(1)' in Eq. \eqref{eq:final_amp_evol} has been omitted for simplicity.

In the absence of the background current ($U_0=0$), we have
\begin{align*}
\mathbf{v_1}&=
\begin{bmatrix}
0,
\dfrac{-\ii k_1 k_2 g}{{\omega}_2 \cosh(k_2 H)},
0
\end{bmatrix}^\dagger,  \\
\mathbf{v_2}&=
\begin{bmatrix}
0;
\dfrac{-\ii k_1 k_2 g}{{\omega}_1 \cosh(k_1 H)};
0
\end{bmatrix}^\dagger,\\
\mathbf{r_j}&=
\begin{bmatrix}
-1,
0,
\dfrac{\ii g}{{\omega}_j}
\end{bmatrix}^\dagger,
\end{align*}
yielding
\begin{align}
\beta_1=\ii \lambda \frac{\omega_1}{2g}\qquad;\qquad\beta_2=\ii \lambda \frac{\omega_2}{2g},
\label{eq:beta_1_2}
\end{align}
\noindent where  $\lambda=\omega_1\omega_2[\sinh{(k_1H)}\sinh{(k_2H)}]^{-1}$. The  subscripts $1$ and $2$ are respectively denoted by $i$ and $r$ in the main text.

\bibliographystyle{apsrev4-2}
\bibliography{apssamp}

 \newcommand{\noop}[1]{}
\begin{thebibliography}{27}%
\makeatletter
\providecommand \@ifxundefined [1]{%
 \@ifx{#1\undefined}
}%
\providecommand \@ifnum [1]{%
 \ifnum #1\expandafter \@firstoftwo
 \else \expandafter \@secondoftwo
 \fi
}%
\providecommand \@ifx [1]{%
 \ifx #1\expandafter \@firstoftwo
 \else \expandafter \@secondoftwo
 \fi
}%
\providecommand \natexlab [1]{#1}%
\providecommand \enquote  [1]{``#1''}%
\providecommand \bibnamefont  [1]{#1}%
\providecommand \bibfnamefont [1]{#1}%
\providecommand \citenamefont [1]{#1}%
\providecommand \href@noop [0]{\@secondoftwo}%
\providecommand \href [0]{\begingroup \@sanitize@url \@href}%
\providecommand \@href[1]{\@@startlink{#1}\@@href}%
\providecommand \@@href[1]{\endgroup#1\@@endlink}%
\providecommand \@sanitize@url [0]{\catcode `\\12\catcode `\$12\catcode
  `\&12\catcode `\#12\catcode `\^12\catcode `\_12\catcode `\%12\relax}%
\providecommand \@@startlink[1]{}%
\providecommand \@@endlink[0]{}%
\providecommand \url  [0]{\begingroup\@sanitize@url \@url }%
\providecommand \@url [1]{\endgroup\@href {#1}{\urlprefix }}%
\providecommand \urlprefix  [0]{URL }%
\providecommand \Eprint [0]{\href }%
\providecommand \doibase [0]{https://doi.org/}%
\providecommand \selectlanguage [0]{\@gobble}%
\providecommand \bibinfo  [0]{\@secondoftwo}%
\providecommand \bibfield  [0]{\@secondoftwo}%
\providecommand \translation [1]{[#1]}%
\providecommand \BibitemOpen [0]{}%
\providecommand \bibitemStop [0]{}%
\providecommand \bibitemNoStop [0]{.\EOS\space}%
\providecommand \EOS [0]{\spacefactor3000\relax}%
\providecommand \BibitemShut  [1]{\csname bibitem#1\endcsname}%
\let\auto@bib@innerbib\@empty
\bibitem [{\citenamefont {Stokes}(1847)}]{stokes1847}%
  \BibitemOpen
  \bibfield  {author} {\bibinfo {author} {\bibfnamefont {G.~G.}\ \bibnamefont
  {Stokes}},\ }\href@noop {} {\bibfield  {journal} {\bibinfo  {journal} {Trans.
  Camb. Philos. Soc.}\ }\textbf {\bibinfo {volume} {8}},\ \bibinfo {pages}
  {441} (\bibinfo {year} {1847})}\BibitemShut {NoStop}%
\bibitem [{\citenamefont {Van~den Bremer}\ and\ \citenamefont
  {Breivik}(2017)}]{van2017stokes}%
  \BibitemOpen
  \bibfield  {author} {\bibinfo {author} {\bibfnamefont {T.~S.}\ \bibnamefont
  {Van~den Bremer}}\ and\ \bibinfo {author} {\bibfnamefont {{\O}.}~\bibnamefont
  {Breivik}},\ }\href@noop {} {\bibfield  {journal} {\bibinfo  {journal}
  {Philos. Trans. R. Soc. A}\ }\textbf {\bibinfo {volume} {376}},\ \bibinfo
  {pages} {20170104} (\bibinfo {year} {2017})}\BibitemShut {NoStop}%
\bibitem [{\citenamefont {Kenyon}(1969)}]{kenyon1969stokes}%
  \BibitemOpen
  \bibfield  {author} {\bibinfo {author} {\bibfnamefont {K.~E.}\ \bibnamefont
  {Kenyon}},\ }\href@noop {} {\bibfield  {journal} {\bibinfo  {journal} {J.
  Geophys. Res.}\ }\textbf {\bibinfo {volume} {74}},\ \bibinfo {pages} {6991}
  (\bibinfo {year} {1969})}\BibitemShut {NoStop}%
\bibitem [{\citenamefont {van Sebille}\ \emph {et~al.}(2020)\citenamefont {van
  Sebille}, \citenamefont {Aliani}, \citenamefont {Law}, \citenamefont
  {Maximenko}, \citenamefont {Alsina}, \citenamefont {Bagaev}, \citenamefont
  {Bergmann}, \citenamefont {Chapron}, \citenamefont {Chubarenko},
  \citenamefont {Cózar}, \citenamefont {Delandmeter}, \citenamefont {Egger},
  \citenamefont {Fox-Kemper}, \citenamefont {Garaba}, \citenamefont
  {Goddijn-Murphy}, \citenamefont {Hardesty}, \citenamefont {Hoffman},
  \citenamefont {Isobe}, \citenamefont {Jongedijk}, \citenamefont {Kaandorp},
  \citenamefont {Khatmullina}, \citenamefont {Koelmans}, \citenamefont
  {Kukulka}, \citenamefont {Laufkötter}, \citenamefont {Lebreton},
  \citenamefont {Lobelle}, \citenamefont {Maes}, \citenamefont
  {Martinez-Vicente}, \citenamefont {Maqueda}, \citenamefont {Poulain-Zarcos},
  \citenamefont {Rodriguez}, \citenamefont {Ryan}, \citenamefont {Shanks},
  \citenamefont {Shim}, \citenamefont {Suaria}, \citenamefont {Thiel},
  \citenamefont {van~den Bremer},\ and\ \citenamefont
  {Wichmann}}]{Sebille2020}%
  \BibitemOpen
  \bibfield  {author} {\bibinfo {author} {\bibfnamefont {E.}~\bibnamefont {van
  Sebille}}, \bibinfo {author} {\bibfnamefont {S.}~\bibnamefont {Aliani}},
  \bibinfo {author} {\bibfnamefont {K.}~\bibnamefont {Law}}, \bibinfo {author}
  {\bibfnamefont {N.}~\bibnamefont {Maximenko}}, \bibinfo {author}
  {\bibfnamefont {J.}~\bibnamefont {Alsina}}, \bibinfo {author} {\bibfnamefont
  {A.}~\bibnamefont {Bagaev}}, \bibinfo {author} {\bibfnamefont
  {M.}~\bibnamefont {Bergmann}}, \bibinfo {author} {\bibfnamefont
  {B.}~\bibnamefont {Chapron}}, \bibinfo {author} {\bibfnamefont
  {I.}~\bibnamefont {Chubarenko}}, \bibinfo {author} {\bibfnamefont
  {A.}~\bibnamefont {Cózar}}, \bibinfo {author} {\bibfnamefont
  {P.}~\bibnamefont {Delandmeter}}, \bibinfo {author} {\bibfnamefont
  {M.}~\bibnamefont {Egger}}, \bibinfo {author} {\bibfnamefont
  {B.}~\bibnamefont {Fox-Kemper}}, \bibinfo {author} {\bibfnamefont
  {S.}~\bibnamefont {Garaba}}, \bibinfo {author} {\bibfnamefont
  {L.}~\bibnamefont {Goddijn-Murphy}}, \bibinfo {author} {\bibfnamefont
  {D.}~\bibnamefont {Hardesty}}, \bibinfo {author} {\bibfnamefont
  {M.}~\bibnamefont {Hoffman}}, \bibinfo {author} {\bibfnamefont
  {A.}~\bibnamefont {Isobe}}, \bibinfo {author} {\bibfnamefont
  {C.}~\bibnamefont {Jongedijk}}, \bibinfo {author} {\bibfnamefont
  {M.}~\bibnamefont {Kaandorp}}, \bibinfo {author} {\bibfnamefont
  {L.}~\bibnamefont {Khatmullina}}, \bibinfo {author} {\bibfnamefont
  {A.}~\bibnamefont {Koelmans}}, \bibinfo {author} {\bibfnamefont
  {T.}~\bibnamefont {Kukulka}}, \bibinfo {author} {\bibfnamefont
  {C.}~\bibnamefont {Laufkötter}}, \bibinfo {author} {\bibfnamefont
  {L.}~\bibnamefont {Lebreton}}, \bibinfo {author} {\bibfnamefont
  {D.}~\bibnamefont {Lobelle}}, \bibinfo {author} {\bibfnamefont
  {C.}~\bibnamefont {Maes}}, \bibinfo {author} {\bibfnamefont {V.}~\bibnamefont
  {Martinez-Vicente}}, \bibinfo {author} {\bibfnamefont {M.}~\bibnamefont
  {Maqueda}}, \bibinfo {author} {\bibfnamefont {M.}~\bibnamefont
  {Poulain-Zarcos}}, \bibinfo {author} {\bibfnamefont {E.}~\bibnamefont
  {Rodriguez}}, \bibinfo {author} {\bibfnamefont {P.}~\bibnamefont {Ryan}},
  \bibinfo {author} {\bibfnamefont {A.}~\bibnamefont {Shanks}}, \bibinfo
  {author} {\bibfnamefont {W.}~\bibnamefont {Shim}}, \bibinfo {author}
  {\bibfnamefont {G.}~\bibnamefont {Suaria}}, \bibinfo {author} {\bibfnamefont
  {M.}~\bibnamefont {Thiel}}, \bibinfo {author} {\bibfnamefont
  {T.}~\bibnamefont {van~den Bremer}},\ and\ \bibinfo {author} {\bibfnamefont
  {D.}~\bibnamefont {Wichmann}},\ }\href@noop {} {\bibfield  {journal}
  {\bibinfo  {journal} {Environ. Res. Lett.}\ } (\bibinfo {year}
  {2020})}\BibitemShut {NoStop}%
\bibitem [{\citenamefont {McWilliams}\ and\ \citenamefont
  {Restrepo}(1999)}]{mcwilliams1999wave}%
  \BibitemOpen
  \bibfield  {author} {\bibinfo {author} {\bibfnamefont {J.}~\bibnamefont
  {McWilliams}}\ and\ \bibinfo {author} {\bibfnamefont {J.~M.}\ \bibnamefont
  {Restrepo}},\ }\href@noop {} {\bibfield  {journal} {\bibinfo  {journal} {J.
  Phys. Oceanogr.}\ }\textbf {\bibinfo {volume} {29}},\ \bibinfo {pages} {2523}
  (\bibinfo {year} {1999})}\BibitemShut {NoStop}%
\bibitem [{\citenamefont {Kumar}\ and\ \citenamefont
  {Feddersen}(2017)}]{kumar2017effect}%
  \BibitemOpen
  \bibfield  {author} {\bibinfo {author} {\bibfnamefont {N.}~\bibnamefont
  {Kumar}}\ and\ \bibinfo {author} {\bibfnamefont {F.}~\bibnamefont
  {Feddersen}},\ }\href@noop {} {\bibfield  {journal} {\bibinfo  {journal} {J.
  Phys. Oceanogr.}\ }\textbf {\bibinfo {volume} {47}},\ \bibinfo {pages} {227}
  (\bibinfo {year} {2017})}\BibitemShut {NoStop}%
\bibitem [{\citenamefont {Clark}(2015)}]{clark2015quantification}%
  \BibitemOpen
  \bibfield  {author} {\bibinfo {author} {\bibfnamefont {M.}~\bibnamefont
  {Clark}},\ }\emph {\bibinfo {title} {Quantification of {S}tokes drift as a
  mechanism for surface oil advection in the {G}ulf of {M}exico during the
  {D}eepwater {H}orizon oil spill}},\ \href@noop {} {Ph.D. thesis},\ \bibinfo
  {school} {The Florida State University} (\bibinfo {year} {2015})\BibitemShut
  {NoStop}%
\bibitem [{\citenamefont {Trinanes}\ \emph {et~al.}(2016)\citenamefont
  {Trinanes}, \citenamefont {Olascoaga}, \citenamefont {Goni}, \citenamefont
  {Maximenko}, \citenamefont {Griffin},\ and\ \citenamefont
  {Hafner}}]{trinanes2016analysis}%
  \BibitemOpen
  \bibfield  {author} {\bibinfo {author} {\bibfnamefont {J.~A.}\ \bibnamefont
  {Trinanes}}, \bibinfo {author} {\bibfnamefont {M.~J.}\ \bibnamefont
  {Olascoaga}}, \bibinfo {author} {\bibfnamefont {G.~J.}\ \bibnamefont {Goni}},
  \bibinfo {author} {\bibfnamefont {N.~A.}\ \bibnamefont {Maximenko}}, \bibinfo
  {author} {\bibfnamefont {D.~A.}\ \bibnamefont {Griffin}},\ and\ \bibinfo
  {author} {\bibfnamefont {J.}~\bibnamefont {Hafner}},\ }\href@noop {}
  {\bibfield  {journal} {\bibinfo  {journal} {J. Oper. Oceanogr.}\ }\textbf
  {\bibinfo {volume} {9}},\ \bibinfo {pages} {126} (\bibinfo {year}
  {2016})}\BibitemShut {NoStop}%
\bibitem [{\citenamefont {Elandt}\ \emph {et~al.}(2014)\citenamefont {Elandt},
  \citenamefont {Shakeri},\ and\ \citenamefont {Alam}}]{elandt2014surface}%
  \BibitemOpen
  \bibfield  {author} {\bibinfo {author} {\bibfnamefont {R.~B.}\ \bibnamefont
  {Elandt}}, \bibinfo {author} {\bibfnamefont {M.}~\bibnamefont {Shakeri}},\
  and\ \bibinfo {author} {\bibfnamefont {M.-R.}\ \bibnamefont {Alam}},\
  }\href@noop {} {\bibfield  {journal} {\bibinfo  {journal} {Phys. Rev. E}\
  }\textbf {\bibinfo {volume} {89}},\ \bibinfo {pages} {023012} (\bibinfo
  {year} {2014})}\BibitemShut {NoStop}%
\bibitem [{\citenamefont {Davies}(1982)}]{davies1982reflection}%
  \BibitemOpen
  \bibfield  {author} {\bibinfo {author} {\bibfnamefont {A.~G.}\ \bibnamefont
  {Davies}},\ }\href@noop {} {\bibfield  {journal} {\bibinfo  {journal} {Dynam.
  Atmos. Oceans}\ }\textbf {\bibinfo {volume} {6}},\ \bibinfo {pages} {207}
  (\bibinfo {year} {1982})}\BibitemShut {NoStop}%
\bibitem [{\citenamefont {Mei}(1985)}]{mei1985resonant}%
  \BibitemOpen
  \bibfield  {author} {\bibinfo {author} {\bibfnamefont {C.~C.}\ \bibnamefont
  {Mei}},\ }\href@noop {} {\bibfield  {journal} {\bibinfo  {journal} {J. Fluid
  Mech.}\ }\textbf {\bibinfo {volume} {152}},\ \bibinfo {pages} {315} (\bibinfo
  {year} {1985})}\BibitemShut {NoStop}%
\bibitem [{\citenamefont {Heathershaw}(1982)}]{heathershaw1982seabed}%
  \BibitemOpen
  \bibfield  {author} {\bibinfo {author} {\bibfnamefont {A.~D.}\ \bibnamefont
  {Heathershaw}},\ }\href@noop {} {\bibfield  {journal} {\bibinfo  {journal}
  {Nature}\ }\textbf {\bibinfo {volume} {296}},\ \bibinfo {pages} {343}
  (\bibinfo {year} {1982})}\BibitemShut {NoStop}%
\bibitem [{\citenamefont {{Ball}}(1964)}]{Ball}%
  \BibitemOpen
  \bibfield  {author} {\bibinfo {author} {\bibfnamefont {F.~K.}\ \bibnamefont
  {{Ball}}},\ }\href@noop {} {\bibfield  {journal} {\bibinfo  {journal} {J.
  Fluid Mech.}\ }\textbf {\bibinfo {volume} {19}},\ \bibinfo {pages} {465}
  (\bibinfo {year} {1964})}\BibitemShut {NoStop}%
\bibitem [{\citenamefont {Heathershaw}\ and\ \citenamefont
  {Davies}(1985)}]{heathershaw1985resonant}%
  \BibitemOpen
  \bibfield  {author} {\bibinfo {author} {\bibfnamefont {A.}~\bibnamefont
  {Heathershaw}}\ and\ \bibinfo {author} {\bibfnamefont {A.}~\bibnamefont
  {Davies}},\ }\href@noop {} {\bibfield  {journal} {\bibinfo  {journal} {Mar.
  Geol.}\ }\textbf {\bibinfo {volume} {62}},\ \bibinfo {pages} {321} (\bibinfo
  {year} {1985})}\BibitemShut {NoStop}%
\bibitem [{\citenamefont {Elgar}\ \emph {et~al.}(2003)\citenamefont {Elgar},
  \citenamefont {Raubenheimer},\ and\ \citenamefont
  {Herbers}}]{elgar2003bragg}%
  \BibitemOpen
  \bibfield  {author} {\bibinfo {author} {\bibfnamefont {S.}~\bibnamefont
  {Elgar}}, \bibinfo {author} {\bibfnamefont {B.}~\bibnamefont
  {Raubenheimer}},\ and\ \bibinfo {author} {\bibfnamefont {T.}~\bibnamefont
  {Herbers}},\ }\href@noop {} {\bibfield  {journal} {\bibinfo  {journal}
  {Geophys. Res. Lett.}\ }\textbf {\bibinfo {volume} {30}} (\bibinfo {year}
  {2003})}\BibitemShut {NoStop}%
\bibitem [{\citenamefont {Couston}\ \emph {et~al.}(2015)\citenamefont
  {Couston}, \citenamefont {Guo}, \citenamefont {Chamanzar},\ and\
  \citenamefont {Alam}}]{reza_fabry}%
  \BibitemOpen
  \bibfield  {author} {\bibinfo {author} {\bibfnamefont {L.-A.}\ \bibnamefont
  {Couston}}, \bibinfo {author} {\bibfnamefont {Q.}~\bibnamefont {Guo}},
  \bibinfo {author} {\bibfnamefont {M.}~\bibnamefont {Chamanzar}},\ and\
  \bibinfo {author} {\bibfnamefont {M.-R.}\ \bibnamefont {Alam}},\ }\href@noop
  {} {\bibfield  {journal} {\bibinfo  {journal} {Phys. Rev. E}\ }\textbf
  {\bibinfo {volume} {92}},\ \bibinfo {pages} {043015} (\bibinfo {year}
  {2015})}\BibitemShut {NoStop}%
\bibitem [{\citenamefont {Kundu}\ \emph {et~al.}(1990)\citenamefont {Kundu},
  \citenamefont {Cohen},\ and\ \citenamefont {Dowling}}]{kundu20fluid}%
  \BibitemOpen
  \bibfield  {author} {\bibinfo {author} {\bibfnamefont {P.~K.}\ \bibnamefont
  {Kundu}}, \bibinfo {author} {\bibfnamefont {I.}~\bibnamefont {Cohen}},\ and\
  \bibinfo {author} {\bibfnamefont {D.}~\bibnamefont {Dowling}},\ }\href@noop
  {} {\bibfield  {journal} {\bibinfo  {journal} {Academic Press}\ }\textbf
  {\bibinfo {volume} {77}},\ \bibinfo {pages} {108} (\bibinfo {year}
  {1990})}\BibitemShut {NoStop}%
\bibitem [{\citenamefont {Constantin}\ and\ \citenamefont
  {Villari}(2008)}]{constantin2008particle}%
  \BibitemOpen
  \bibfield  {author} {\bibinfo {author} {\bibfnamefont {A.}~\bibnamefont
  {Constantin}}\ and\ \bibinfo {author} {\bibfnamefont {G.}~\bibnamefont
  {Villari}},\ }\href@noop {} {\bibfield  {journal} {\bibinfo  {journal} {J.
  Math. Fluid Mech.}\ }\textbf {\bibinfo {volume} {10}},\ \bibinfo {pages} {1}
  (\bibinfo {year} {2008})}\BibitemShut {NoStop}%
\bibitem [{\citenamefont {Ursell}(1953)}]{ursell1953long}%
  \BibitemOpen
  \bibfield  {author} {\bibinfo {author} {\bibfnamefont {F.}~\bibnamefont
  {Ursell}},\ }in\ \href@noop {} {\emph {\bibinfo {booktitle} {Math. Proc.
  Camb. Philos. Soc}}},\ Vol.~\bibinfo {volume} {49}\ (\bibinfo {organization}
  {Cambridge University Press},\ \bibinfo {year} {1953})\ pp.\ \bibinfo {pages}
  {685--694}\BibitemShut {NoStop}%
\bibitem [{\citenamefont {Liu}\ and\ \citenamefont
  {Yue}(1998)}]{liu1998generalized}%
  \BibitemOpen
  \bibfield  {author} {\bibinfo {author} {\bibfnamefont {Y.}~\bibnamefont
  {Liu}}\ and\ \bibinfo {author} {\bibfnamefont {D.~K.~P.}\ \bibnamefont
  {Yue}},\ }\href@noop {} {\bibfield  {journal} {\bibinfo  {journal} {J. Fluid
  Mech.}\ }\textbf {\bibinfo {volume} {356}},\ \bibinfo {pages} {297} (\bibinfo
  {year} {1998})}\BibitemShut {NoStop}%
\bibitem [{\citenamefont {Davies}\ and\ \citenamefont
  {Heathershaw}(1984)}]{davies1984surface}%
  \BibitemOpen
  \bibfield  {author} {\bibinfo {author} {\bibfnamefont {A.~G.}\ \bibnamefont
  {Davies}}\ and\ \bibinfo {author} {\bibfnamefont {A.~D.}\ \bibnamefont
  {Heathershaw}},\ }\href@noop {} {\bibfield  {journal} {\bibinfo  {journal}
  {J. Fluid Mech.}\ }\textbf {\bibinfo {volume} {144}},\ \bibinfo {pages} {419}
  (\bibinfo {year} {1984})}\BibitemShut {NoStop}%
\bibitem [{\citenamefont {Dommermuth}\ and\ \citenamefont
  {Yue}(1987)}]{dommermuth1987high}%
  \BibitemOpen
  \bibfield  {author} {\bibinfo {author} {\bibfnamefont {D.~G.}\ \bibnamefont
  {Dommermuth}}\ and\ \bibinfo {author} {\bibfnamefont {D.~K.~P.}\ \bibnamefont
  {Yue}},\ }\href@noop {} {\bibfield  {journal} {\bibinfo  {journal} {J. Fluid
  Mech.}\ }\textbf {\bibinfo {volume} {184}},\ \bibinfo {pages} {267} (\bibinfo
  {year} {1987})}\BibitemShut {NoStop}%
\bibitem [{\citenamefont {Raj}\ and\ \citenamefont
  {Guha}(2019)}]{raj_guha_2019}%
  \BibitemOpen
  \bibfield  {author} {\bibinfo {author} {\bibfnamefont {R.}~\bibnamefont
  {Raj}}\ and\ \bibinfo {author} {\bibfnamefont {A.}~\bibnamefont {Guha}},\
  }\href@noop {} {\bibfield  {journal} {\bibinfo  {journal} {J. Fluid Mech.}\
  }\textbf {\bibinfo {volume} {867}},\ \bibinfo {pages} {482–515} (\bibinfo
  {year} {2019})}\BibitemShut {NoStop}%
\bibitem [{\citenamefont {Zakharov}(1968)}]{zakharov1968stability}%
  \BibitemOpen
  \bibfield  {author} {\bibinfo {author} {\bibfnamefont {V.~E.}\ \bibnamefont
  {Zakharov}},\ }\href@noop {} {\bibfield  {journal} {\bibinfo  {journal} {J.
  Appl. Mech. Tech. Phys.}\ }\textbf {\bibinfo {volume} {9}},\ \bibinfo {pages}
  {190} (\bibinfo {year} {1968})}\BibitemShut {NoStop}%
\bibitem [{\citenamefont {Kirby}(1988)}]{kirby1988current}%
  \BibitemOpen
  \bibfield  {author} {\bibinfo {author} {\bibfnamefont {J.~T.}\ \bibnamefont
  {Kirby}},\ }\href@noop {} {\bibfield  {journal} {\bibinfo  {journal} {J.
  Fluid Mech.}\ }\textbf {\bibinfo {volume} {186}},\ \bibinfo {pages} {501}
  (\bibinfo {year} {1988})}\BibitemShut {NoStop}%
\bibitem [{\citenamefont {Dohan}(2017)}]{dohan2017ocean}%
  \BibitemOpen
  \bibfield  {author} {\bibinfo {author} {\bibfnamefont {K.}~\bibnamefont
  {Dohan}},\ }\href@noop {} {\bibfield  {journal} {\bibinfo  {journal} {J.
  Geophys. Res. Oceans}\ }\textbf {\bibinfo {volume} {122}},\ \bibinfo {pages}
  {2647} (\bibinfo {year} {2017})}\BibitemShut {NoStop}%
\bibitem [{\citenamefont {Chubarenko}\ \emph {et~al.}(2018)\citenamefont
  {Chubarenko}, \citenamefont {Esiukova}, \citenamefont {Bagaev}, \citenamefont
  {Isachenko}, \citenamefont {Demchenko}, \citenamefont {Zobkov}, \citenamefont
  {Efimova}, \citenamefont {Bagaeva},\ and\ \citenamefont
  {Khatmullina}}]{chubarenko2018behavior}%
  \BibitemOpen
  \bibfield  {author} {\bibinfo {author} {\bibfnamefont {I.}~\bibnamefont
  {Chubarenko}}, \bibinfo {author} {\bibfnamefont {E.}~\bibnamefont
  {Esiukova}}, \bibinfo {author} {\bibfnamefont {A.}~\bibnamefont {Bagaev}},
  \bibinfo {author} {\bibfnamefont {I.}~\bibnamefont {Isachenko}}, \bibinfo
  {author} {\bibfnamefont {N.}~\bibnamefont {Demchenko}}, \bibinfo {author}
  {\bibfnamefont {M.}~\bibnamefont {Zobkov}}, \bibinfo {author} {\bibfnamefont
  {I.}~\bibnamefont {Efimova}}, \bibinfo {author} {\bibfnamefont
  {M.}~\bibnamefont {Bagaeva}},\ and\ \bibinfo {author} {\bibfnamefont
  {L.}~\bibnamefont {Khatmullina}},\ }in\ \href@noop {} {\emph {\bibinfo
  {booktitle} {Microplastic Contamination in Aquatic Environments}}}\ (\bibinfo
   {publisher} {Elsevier},\ \bibinfo {year} {2018})\ pp.\ \bibinfo {pages}
  {175--223}\BibitemShut {NoStop}%
\end{thebibliography}%

\end{document}